\documentclass[letterpaper,11pt]{article}
\pdfoutput=1

\usepackage{xspace}
\usepackage{slashed}
\usepackage{array,multirow}
\usepackage{graphicx}
\usepackage{graphics}
\usepackage{epsfig}
\usepackage{amsfonts}
\usepackage{amsmath}
\usepackage{amssymb, bbold}
\usepackage{dsfont}
\usepackage{color}
\usepackage{xcolor}
\usepackage{cite}
\usepackage{pdflscape}
\usepackage{pifont}
\usepackage{pgfplots}
\usepackage{tikz} 
\usepackage{pgfplotstable}
\usepackage{enumitem}
\usepackage{cleveref}

\usepackage{bm}  %

\newcommand{\be}{\begin{equation}}
\newcommand{\ee}{\end{equation}}
\newcommand{\bea}{\begin{eqnarray}}
\newcommand{\eea}{\end{eqnarray}}
\newcommand{\ba}{\begin{array}}
\newcommand{\ea}{\end{array}}
\newcommand{\eps}{\epsilon}
\renewcommand\({\left(}
\renewcommand\){\right)}
\renewcommand\[{\left[}

\newcommand{\dd}{{\rm d}}
\newcommand{\e}{{\rm e}}

\def\nn{\nonumber}

\renewcommand{\Im}{{\rm Im}}

\def\slashchar#1{\setbox0=\hbox{$#1$}           
  \dimen0=\wd0                                    
  \setbox1=\hbox{/} \dimen1=\wd1                  
  \ifdim\dimen0>\dimen1                           
    \rlap{\hbox to \dimen0{\hfil/\hfil}}            
    #1                                             
  \else                                          
    \rlap{\hbox to \dimen1{\hfil$#1$\hfil}}        
    /                                           
 \fi}

\usepackage{tikz}
\usetikzlibrary{arrows,shapes}
\usetikzlibrary{trees}
\usetikzlibrary{matrix,arrows} 				
\usetikzlibrary{positioning}				
\usetikzlibrary{calc,through}				
\usetikzlibrary{decorations.pathreplacing}  
\usepackage{pgffor}							

\usetikzlibrary{decorations.pathmorphing}	
\usetikzlibrary{decorations.markings}
\tikzset{
    vector/.style={decorate, decoration={snake}, draw},
	provector/.style={decorate, decoration={snake,amplitude=2.5pt}, draw},
	antivector/.style={decorate, decoration={snake,amplitude=-2.5pt}, draw},
    fermion/.style={draw=black, postaction={decorate},
        decoration={markings,mark=at position .55 with {\arrow[draw=black]{>}}}},
    fermionbar/.style={draw=black, postaction={decorate},
        decoration={markings,mark=at position .55 with {\arrow[draw=black]{<}}}},
    fermionnoarrow/.style={draw=black},
    gluon/.style={decorate, draw=black,
        decoration={coil,amplitude=4pt, segment length=5pt}},
    scalar/.style={dashed,draw=black, postaction={decorate},
        decoration={markings,mark=at position .55 with {\arrow[draw=black]{>}}}},
    scalarbar/.style={dashed,draw=black, postaction={decorate},
        decoration={markings,mark=at position .55 with {\arrow[draw=black]{<}}}},
    scalarnoarrow/.style={dashed,draw=black},
    electron/.style={draw=black, postaction={decorate},
        decoration={markings,mark=at position .55 with {\arrow[draw=black]{>}}}},
	bigvector/.style={decorate, decoration={snake,amplitude=4pt}, draw},
}

\begin{document}

\begin{titlepage}

\begin{flushright}
Nikhef-2018-056
\end{flushright}

\vspace{2.0cm}

\begin{center}
{\LARGE  \bf 
The role of leptons in electroweak baryogenesis 
}

\vspace{2.4cm}

{\large \bf Jordy de Vries$^{a,b}$, Marieke Postma$^c$, Jorinde van de Vis$^c$}
\vspace{0.5cm}

\vspace{0.25cm}

\vspace{0.25cm}
{\large 
$^a$ 
{\it 
Amherst Center for Fundamental Interactions, Department of Physics, \\
University of Massachusetts, Amherst, MA 01003}}

\vspace{0.25cm}
{\large 
$^b$ 
{\it 
RIKEN BNL Research Center, Brookhaven National Laboratory, \\Upton, NY 11973-5000}}

{\large 
$^c$ 
{\it Nikhef, Theory Group, Science Park 105, 1098 XG, Amsterdam, The Netherlands}}

\end{center}

\vspace{1.5cm}

\begin{abstract}
We investigate the role of leptons in electroweak baryogenesis by studying a relatively simple framework inspired by effective field theory 
that satisfies all Sakharov conditions. In particular, we study the effectiveness of CP-violating source terms induced by dimension-six Yukawa interactions
for quarks and charged leptons. Despite the relatively small Yukawa coupling, CP-violating source terms involving taus are quite effective
and can account for the observed matter-antimatter asymmetry. We obtain analytical and numerical expressions for the total baryon asymmetry, the former providing important insight into what makes lepton CP violation relatively effective compared to quark CP violation. Leptons also play an important role if the CP-violating source involves top quarks. While the tau Yukawa coupling in the Standard Model is small, it significantly enhances the baryon asymmetry by transferring the chiral asymmetry in quarks, which is washed out by strong sphalerons, to a chiral asymmetry in leptons. We conclude that leptons should not be ignored even if CP violation is limited to the quark sector. 
The role of leptons can be further increased in scenarios of new physics with additional chiral-symmetry-breaking interactions between quarks and leptons, as can happen in models with additional Higgs bosons or leptoquarks. Finally, we study CP-violating dimension-six Yukawa interactions for lighter quarks and leptons but conclude that these  lead to too small baryon asymmetries.

\end{abstract}

\vfill
\end{titlepage}


\section{Introduction}\label{sec:intro}

Understanding the prevalence of matter over antimatter in our universe is one of the great challenges in particle physics and cosmology. The baryon asymmetry can be extracted from the Planck data on the cosmic microwave background \cite{Ade:2015xua}
\be
Y_B \equiv \frac{n_b}{s} = (8.50 \pm 0.11) \times 10^{-11} \, ,
\label{BAU}
\ee
with $n_b$ and $s$ the baryon number and entropy density respectively.  To dynamically explain this number requires satisfying the three Sakharov conditions \cite{Sakharov:1967dj}: 1) baryon number violation, 2) charge (C) and charge-parity (CP) violation, and 3) out-of-equilibrium dynamics. The standard model (SM) only fulfills the first one -- electroweak sphaleron transitions active at high temperatures violate baryon number -- and physics beyond the standard model is needed to explain the baryon asymmetry in the universe.

Of the existing baryogenesis theories, electroweak baryogenesis (EWBG) is particularly interesting as it is linked to electroweak scale physics and can be tested in experiments. The (minimal) beyond-the-Standard Model (BSM) ingredients in this scenario are new sources of CP violation, and an extended scalar sector that can give rise to a first-order electroweak phase transition that provides the necessary out-of-equilibrium dynamics.  Both ingredients can be probed by the large hadron collider (LHC), for instance via searches for new scalars \cite{Arhrib:2013oia, Chen:2013rba, Chang:2017ynj, CMS:2016tgd}, precision Higgs studies \cite{Englert:2014uua, Brivio:2017vri}, and CP-odd collider observables \cite{Han:2009ra, Boudjema:2015nda, Ellis:2015dha, Askew:2015mda, Demartin:2015uha}.  Typically, the best constraints on new sources of CP violation, however, come from electric dipole moment measurements \cite{Chupp:2017rkp, Balazs:2016yvi, deVries:2017ncy}.  Gravitational waves produced during a first-order electroweak phase transition may be measured by LISA or other future gravitational wave detectors \cite{Audley:2017drz, Caprini:2015zlo}.

The mechanism of EWBG is, in a nutshell, as follows. The first-order  electroweak phase transition proceeds via the formation of bubbles of broken Higgs vacuum, which subsequently expand to eventually fill up all of space.  The quarks and leptons in the plasma collide with the walls of the expanding bubbles. If these interactions violate CP, the transmission and reflection coefficients are different for particles and antiparticles. The net result is that an overdensity of left-handed particles over antiparticles builds up in front of the bubble wall.  The (B+L)-violating  electroweak sphaleron transitions only act on electroweak doublets, and transform this ``chiral asymmetry'' into a net baryon number in front of the bubble.  The produced baryons are then swept up by the expanding bubble. Inside the bubble the baryon number is preserved as the electroweak sphaleron processes are strongly suppressed in the broken vacuum. For reviews of EWBG, see for example Refs.~\cite{Morrissey:2012db, Cline:2006ts, RevModPhys.71.1463, White:2016nbo}.

A large number of SM extensions that could lead to successful EWBG have been proposed. In principle, a detailed phenomenological study of each individual model is required to test the feasibility of the proposed explanation of the matter-antimatter asymmetry. To avoid such a cumbersome exercise it was proposed in Ref.~\cite{Balazs:2016yvi} to test EWBG, or at least a large class of EWBG models, in a single framework based on the Standard Model Effective Field Theory (SM-EFT). Unfortunately, a detailed study in Ref.~\cite{deVries:2017ncy} concluded that EWBG cannot be fully studied within the SM-EFT framework.  The main reason is that the SM-EFT breaks down in the scalar sector, because new \emph{light} degrees of freedom are necessary to obtain a strong first-order phase transition. This breakdown is communicated to the CP-violating (CPV) sector of the theory via the equation of motion of the Higgs field, which is used to construct the basis set of SM-EFT operators. The result is that there is no separation of scales and higher-dimensional operators cannot be neglected, thus invalidating the SM-EFT approach. In principle, the breakdown of EFT methods occurs in the scalar sector, which thus requires new light degrees of freedom, while the CPV sector can potentially still be described by effective operators that can now also contain the new degrees of freedom.

 In this work, we mainly focus on the CPV dynamics and avoid the issue of the first-order phase transition by describing the bubble-wall profile in terms of a phenomenological tanh-function which provides a reasonable description of actual solutions. 
We describe the required additional CPV by effective dimension-six CPV operators containing SM fields only. As mentioned, in principle the CPV dynamics could arise from effective operators involving any new fields that play a role in the phase transition \cite{Vaskonen:2016yiu}, but as these are difficult to probe in experiments we ignore such interactions for now. In particular, we focus on effective dimension-six Yukawa interactions of various quarks and leptons as these are representative for popular classes of BSM models such as multi-Higgs models. 
Similar studies in the literature have focussed on CPV in the top-quark sector \cite{Huber:2006ri,Balazs:2016yvi,  deVries:2017ncy}, as the large top Yukawa coupling maximizes the CPV source term in the transport equations that describe the dynamics of the particle number densities.  However, taking into account the very recent new constraint on CPV operators from an improved measurement of the electric dipole moment of the electron \cite{Andreev:2018ayy}, the `top-source' scenario gives an asymmetry that is about two orders of magnitude too small   to explain the baryon asymmetry \cite{deVries:2017ncy}.

The small value of the baryon asymmetry in the top-source scenario has prompted our current study of the general features of the solutions to the transport equations, to find ways to boost the asymmetry. The reasons for the inefficiency of the top-source scenario are threefold. First, the diffusion of the chiral asymmetry into the symmetric phase is not efficient for the strongly-interacting quarks \cite{Joyce:1994bi}. Second, EDM measurements put strong constraints on CPV in the top sector \cite{Cirigliano:2016nyn} such that the strength of the CPV source term is limited. And thirdly, the  washout of the produced chiral asymmetry is significant for (top) quarks,
as the top Yukawa and especially the strong sphaleron interactions effectively wash out the chiral asymmetry in the symmetric phase, except for regions very close to the bubble wall \cite{Giudice:1993bb, Tulin:2011wi}.
These problems can potentially be overcome by looking at CPV source terms involving lighter fermions. For instance, EDM limits are less stringent for bottom quarks. While the CPV source term for bottom quarks is suppressed by the smaller Yukawa coupling, the washout rate due to Higgs interactions is suppressed accordingly. As such, the total baryon asymmetry is not a simple function of the size of the Yukawa coupling. 

For leptons there can be even more advantages even though they have been neglected in many studies of EWBG. While the CPV source term is suppressed, leptons diffuse into the plasma much more easily  \cite{Joyce:1994bi}, the EDM limits are less stringent for muon and tau CPV interactions \cite{Brod:2013cka}, and the washout rate is less effective because leptons do not interact via strong sphalerons. Already in Refs. \cite{Joyce:1994bi, Joyce:1994zn, Chung:2009cb, Chiang:2016vgf, Guo:2016ixx } the effect of leptons on EWBG was studied in a range of scenarios in which the lepton Yukawa coupling is enhanced significantly with respect to the SM values. We will show that even with the small SM Yukawa interactions, including leptons can dramatically change the baryon asymmetry in models with CPV source terms involving top quarks. In addition, we show that leptonic CPV source terms can be very efficient in producing the baryon asymmetry of the universe. Scenarios with a CPV leptonic source are also great diagnostic tools as the set of transport equations is relatively simple. We use this scenario to understand the parametric dependence of the baryon asymmetry on bubble wall parameters and the size of Yukawa couplings. This study confirms that the baryon asymmetry is not a simple function of the size of the Yukawa coupling. And while we focus on a particular set-up, the importance of leptons and the general mechanisms at play are more general.

As we will show, the role of leptons in EWBG depends on the effectiveness of the exchange of the chiral asymmetry between quarks and leptons. In the SM this exchange is not very efficient because of the small lepton Yukawa interactions. In various BSM models there can be more efficient mechanisms, for example via the exchange of additional scalar fields. Such mechanisms can strongly boost the baryon asymmetry if the CPV source term is located in the quark sector by transferring the chiral asymmetry into the lepton sector, where it diffuses faster and suffers from less washout. Conversely, it can suppress  the baryon asymmetry if the CPV source term is located in the lepton sector. We model this phenomenon by adding effective dimension-six quark-lepton interactions; this set-up qualitatively explains features of various models studied in recent literature \cite{Chiang:2016vgf, Guo:2016ixx}. We also comment on the effect of possible new BSM quark-quark couplings, which may likewise boost the baryon asymmetry for a source located in the quark sector, as this coupling affects and limits the washout from strong sphaleron interactions.

This paper is organized as follows.  In the next section we introduce the set-up and briefly describe the first-order phase transition and bubble profile, the CPV dimension-six operators and the EDM and LHC constraints. We also list the transport equations that track the number densities of the particles in the plasma. We then identify the factors that suppress the value of the baryon asymmetry in the top-source scenario and motivate the importance of leptons. We also introduce a new top-lepton interaction. We start with a discussion of the tau-source scenario in \cref{s:tau}. The lepton sector almost completely decouples from the quark sector, and the baryon asymmetry can be computed analytically in this limit to good accuracy. We identify the important length scales, and discuss the physics and mechanisms at play.  In \cref{sec:quarks} we give the results for the top-source scenario.  The equations to solve are more intricate, but the outcome can be understood qualitavitely. We also comment on the viability of a bottom-source scenario and briefly discuss even lighter fermions.  Finally, in \cref{s:extra} we describe the effects of a new tau-top coupling on both the tau- and top-source scenario.  We end with a discussion in \cref{s:conclusions}.


\section{Set-up and methods}

In this section we present our set-up, and briefly review the ingredients that go into the calculation of the baryon asymmetry.

\subsection{First-order phase transition}\label{sec:FOPS}

Sakharov's out-of-equilibrium condition is not satisfied in the SM, where the electroweak phase transition is a cross-over \cite{Gurtler:1997hr, Laine:1998jb, Aoki:1999fi, Csikor:1998eu}. There are many BSM models that modify the Higgs sector such that the phase transition becomes first order.  Well studied examples are the $\mathbb{Z}_2$-symmetric Higgs-singlet model \cite{Espinosa:1993bs,Espinosa:2007qk,Barger:2007im,Espinosa:2008kw, Espinosa:2011ax,Cline:2012hg} and two-Higgs doublet models \cite{Bochkarev:1990fx,Turok:1991uc,Davies:1994id ,Cline:1996mga,Cline:2011mm,Dorsch:2016nrg, Andersen:2017ika,Gorda:2018hvi}.  In our previous work \cite{deVries:2017ncy}, folowing up on earlier work by \cite{Damgaard:2015con }, we argued that a first-order electroweak phase transition cannot be described in a systematic EFT expansion and explicit light degrees of freedom must be introduced. 
This unfortunately prohibits a model-independent approach, and it is necessary to pick a specific BSM model to implement the phase transition.

Once a BSM model is chosen, one can solve the tunnelling equations of motion at the nucleation temperature $T_N$ to find the so-called bounce solution $\phi_b$ \cite{PhysRevD.15.2929}. We  parameterize the Higgs doublet as $H= (H^+ , H^0)^T$ and in the bubble background $\langle H_0\rangle = \frac1{\sqrt{2}} \phi_b(z)$.
In this work, we will use the commonly used parametrization \cite{John:1998ip}
\be
\phi_b(z) = \frac{v_N}{2} \left (1 + \tanh{\frac{z}{L_w}}\right ),
\label{tanh}
\ee
which provides a reasonable description of the bubble-wall profile in many models. More complicated profiles can be studied in similar fashion. We do not expect our findings in this work to change significantly if more complicated profiles are applied.
The ansatz assumes the planar wall approximation, in which curvature effects are neglected and all functions only depend on the distance $z$ from the center of the wall in the bubble-wall rest frame. The broken phase extends to $z\rightarrow \infty$ and the symmetric phase to $z \rightarrow -\infty$.  Further, $v_N= v(T_N)$ is the vacuum expectation value of the Higgs field in the broken phase at the nucleation temperature $T_N$, and $L_w$ the bubble width.  In principle, these parameters should be determined by fitting the bounce solution of a BSM model to \cref{tanh}. However, our prime interest is not the phase transition itself but the comparison of different sources of CP violation.  We work with benchmark parameters $T_N =88 \, \text{GeV} $, $v_N= 152 \, \text{GeV} $, and $L_w = 0.11 \, \text{GeV}^{-1}$ based on an explicit solution in our earlier work \cite{deVries:2017ncy}. For the velocity of the bubble wall we take the benchmark value $v_w = 0.05$. We will investigate how the produced asymmetry depends on these parameters.
			
\subsection{Source of CP violation}

The CP-violating phase of the CKM-matrix of the Standard Model is not sufficient to explain the observed baryon asymmetry \cite{Gavela:1993ts, Gavela:1994dt, Huet:1994jb, Brauner:2012gu} and BSM physics should provide an extra source of CP violation. In this work, we describe this CP violation with effective dimension-six operators. That is, we assume that apart from a modified scalar sector that ensures a first-order phase transition, other BSM degrees of freedom are sufficiently heavy and can be integrated out leading to effective operators. In particular, we consider the flavor-diagonal CPV dimension-six operators
\begin{equation}
  \mathcal L_6 = -i  \left[ \bar Q_L  \tilde Y_U \tilde H  \,u_R\  + \, \bar Q_L  \tilde Y_D  H  \,d_R\,   + \, \bar L_L  \tilde Y_L  H  \,e_R\right](H^\dagger H) + \mathrm{h.c.}\,,
  \label{L6}
\end{equation}
in terms of the Higgs doublet $H$ (with $\tilde H^a =\epsilon^{ab} H^{b*}$), the left-handed quark and lepton $SU(2)$ doublets $Q_L$ and $L_L$, and the right-handed up, down, and lepton singlets $u_R$, $d_R$, and $e_R$. We have suppressed generation indices and consider the $3\times3$ matrices of Wilson coefficients $\tilde c_{U,D,L}$ to be diagonal and real (such that the operators are purely CP violating). The extension to include flavor-changing operators can be made straightforwardly. We assume the dimension-six Yukawa couplings to be proportional to the SM Yukawa couplings $y_f/\sqrt{2} = m_f/v$ ($v\simeq 246$ GeV is the zero-temperature vev) as is the case in many BSM scenarios and suggested by minimal flavor violation
\begin{eqnarray}
  \tilde Y_{U} = \mathrm{diag}(y_u \tilde c_u,\,y_c\tilde c_c,\,y_t \tilde c_t) \,, \;\;
  \tilde Y_{D} = \mathrm{diag}(y_d \tilde c_d,\,y_s \tilde c_s,\,y_b \tilde c_b) \,,\;\;
  \tilde Y_{L} = \mathrm{diag}(y_e \tilde c_e,\,y_\mu \tilde c_\mu,\,y_\tau \tilde c_\tau) \,.
\end{eqnarray}
Finally, we write
\begin{equation}
\tilde c_f = \frac{s_f}{\Lambda_f^2}\,,
\end{equation}
where $s_f = \pm1$ is chosen to obtain a net number of baryons (rather than antibaryons) and $\Lambda_f$ is the associated scale of new physics where the EFT breaks down.
 We stress that these operators are just a subset of dimension-six SM-EFT CPV operators that can be constructed. The above operators are particularly efficient in generating a baryon asymmetry as they give rise to an effective CPV mass term during the phase transition. In addition to SM-EFT operators, in principle there can be CPV operators that include the unspecified light scalar degrees of freedom, which can only be included if we consider a specific UV completion of the scalar sector. We focus instead on the operators in \cref{L6} as these can readily be constrained by EDM experiments.

The transport equations that track the number densities of the plasma particles are computed in the Closed Time Path (CTP) formalism using the methods in Ref.~\cite{Lee:2004we}. During the electroweak phase transition in the presence of a bubble, the effective fermion mass is spacetime-dependent.  We split  $m_f = \bar m_{f}(\phi_b,T) +m_{f,T}(T)$, with $m_{f,T}$ the usual finite-temperature mass and 
\be
\bar m_{f} = \frac{y_f}{\sqrt{2}}\phi_b \(1 + i s_f \frac{\phi_b^2}{2\Lambda_{f}^2}+...\),
\label{mbar2}
\ee where the ellipses denote the finite temperature corrections to the dimension-six operators in \cref{L6}, which can be neglected in the high-temperature expansion \cite{deVries:2017ncy}.  In the bubble background the phase of the mass matrix cannot be rotated away globally. The effective mass differs for left- and right-handed particles and antiparticles, and consequently they scatter differently off the bubble wall. This gives a source term in the transport equations that drives the chiral asymmetry.

There are different appraoches for calculating the source term. The vev-insertion approximation (VIA)  method of \cite{Lee:2004we} treats the field-dependent mass \cref{mbar2} as a perturbation.  The source arises at first order in the derivative expansion, and depends crucially on the thermal corrections, as it is absent for vanishing thermal width of the fermion. Explicitly\footnote{We neglect the effects of collective plasma excitations, i.e. of the hole modes.}
\be
S_f= \frac{v_w N_c}{\pi^2} \Im (\bar m_f' \bar m_f^*) J_f(T)
= s_f \frac{v_w N_c}{\pi^2} \frac{y_f^2 \phi_b^3 \phi_b'}{2\Lambda_f^2}  J_f(T)\,
\label{Sf2}
\ee
with $N_c$ the number of colors (we set $N_c=1$ if $f$ is a lepton), and $J_f(T)$ an integral expression that depends on the thermal masses of the right- and left-handed fermion and on their thermal widths \cite{Lee:2004we}. The prime denotes derivation with respect to $z$, the distance to the center of the bubble wall.  Although the higher order corrections are not explicitly known, on dimensional grounds the perturbative expansion in VIA is in terms of the parameter $\eps_f = \bar m_{f}^2/T^2$, with $\bar m_{f}$ the zero temperature fermion mass \cref{mbar2}. For the top quark $\eps_t = {\mathcal O}(1)$ and the expansion probably breaks down, whereas for all other SM fermions $\eps_f \ll 1$.

An alternative approach is the semi-classical method of \cite{Cline:2017jvp,Joyce:1994fu,Cline:2000nw}, which uses the WKB expansion, and is also valid for large Yukawa couplings. The source term does not depend on the thermal corrections. The expression found in \cite{Cline:2000nw} (multiplied by $T^3/6$ to match our conventions) is
\be
S_f^{\rm WKB} =\frac{c v_w N_c D_f}{12} (|\bar m_f|^2 \theta_f')''
\ee
with $c = {\cal O}(1)$, $D_f$ the diffusion constant, and $\theta = \arg(\bar m_f)$. Since it is third order in the derivative expansion it is a factor $(L_w T_N)^{-2} \sim 10^{-2}$ suppressed with respect to the source in \cref{Sf2},  where we used the benchmark values listed in \cref{A:BM}.  Including all numerical factors negates this suppression: for our benchmark values we find that the semi-anytical and VIA source terms are of the same order of magnitude $S_f^{\rm WKB}/S_f = 0.2 -5 $, with the larger values obtained for leptons (which have larger diffusion constants). For definiteness, in this paper we will use the source term in \cref{Sf2} (and also the CP-conserving relaxation rate) derived in VIA for all fermions.  The qualitative results on the role of the leptons in EWBG will not depend on this, and the results can straightforwardly be adapted to different source terms.

\subsection{Experimental constraints on CP-violating dimension-six operators}
\label{sec:EDM}
The dimension-six operators in \cref{L6} induce CPV couplings between Higgs bosons and fermions. In the SM once we perform a field redefinition to ensure real and diagonal fermion masses, the Higgs-fermion interactions become CP conserving and flavor diagonal. In the presence of the dimension-six operators, the total Higgs-fermion interactions are still flavor diagonal (by construction) but no longer CP conserving. We obtain
\begin{equation}
\mathcal L_h = -\frac{m_f}{v}\bar f\,f\,h - \frac{s_f m_f}{\Lambda_f^2}\,\bar f\,i\gamma^5 f\,\left(v h + \frac{3}{2}h^2 + \frac{1}{2}\frac{h^3}{v}\right)\,,
\end{equation}
in terms of the real fermion mass $m_f$ and $f$ sums over all quarks and charged leptons. EDM experiments can constrain some of the CPV $\bar f\,i\gamma^5 t\,h$ interactions, depending on the fermion $f$. As we will argue, for EWBG purposes the only relevant interactions are those involving the top, bottom, and tau and we mainly discuss these. Couplings to lighter fermions are too weak to create sufficient baryon asymmetry.

The CPV couplings involving top, bottom, and tau fermions all induce a contribution to the electron EDM via two-loop Barr-Zee diagrams \cite{Brod:2013cka,Chien:2015xha, Brod:2018pli}. Their contributions are given by
\begin{equation}\label{deA}
\frac{d_e}{e} = - \frac{8 \alpha_{em}}{(4\pi)^3} m_e\left[ N_c Q_t^2\, g(x_t)\,\frac{s_t}{\Lambda_t^2} + N_c Q_b^2 \, g(x_b)\,\frac{s_b}{\Lambda_b^2} + Q_\tau^2 \, g(x_\tau)\,\frac{s_\tau}{\Lambda_\tau^2}    \right]\, ,
\end{equation}
where $Q_f$ denotes the fermion charge in units of $e$, $x_f = m_f^2/m_h^2$, and $g(x_f)$ the two-loop function 
\begin{equation}
g(x_f) = \frac{x_f}{2}\int_0^1 dx\,\frac{1}{x(1-x)-x_f}\log\left(\frac{x(1-x)}{x_f}\right)\,.
\end{equation}
Numerically we have $g(x_t) \simeq 1.4$, $g(x_b) \simeq 2.7 \cdot 10^{-2}$, $g(x_\tau) \simeq 7.7 \cdot 10^{-3}$, and for lighter fermions  the function roughly scales as $g(x_f) \sim x_f \log x_f$. In these expressions we have for simplicity neglected QCD renormalization-group effects that mild affect the constraints for quarks \cite{Brod:2013cka,Chien:2015xha, Brod:2018pli}.

The ACME experiment has recently improved the constraint on the electron EDM to $d_e \leq 1.1 \times 10^{-29}\,e\,\mathrm{cm}$ at $90\%$ c.l. \cite{Andreev:2018ayy}. Inserting this into the above expression shows that this sets a strong constraint on the CPV top-Higgs coupling for which we obtain $\Lambda_t \geq 7.1 $ TeV. For the bottom and charm we find $\Lambda_{b} \geq 0.49 $ TeV and $\Lambda_{c} \geq 0.41$ TeV and for the tau $\Lambda_{\tau} \geq 0.16$ TeV. For the lighter fermions no meaningful constraint can be set as the the limit on $\Lambda_f$ is lower than the electroweak scale. The tau coupling can in principle also be constrained by the limit on the tau EDM. However, while the contribution to the tau EDM from CPV tau-couplings is about a factor $\mathcal O(10^6)$ larger than the contribution to the eEDM, the experimental limit on the tau EDM is roughly a factor $\mathcal O(10^{11})$ weaker \cite{Inami:2002ah} and no significant constraints are obtained. The story is similar for the CPV $\mu$-Higgs coupling and no significant constraint can be set. A CPV $e$-Higgs coupling, however, would lead to a large eEDM and we get a limit $\Lambda_e \geq 5.7$ TeV \cite{Altmannshofer:2015qra}.

Additional constraints can be set by using experimental limits on hadronic EDMs. In this case, the analysis is more complicated and requires apart from several additional one- and two-loop diagrams also renormalization-group evolution factors and hadronic and nuclear matrix elements. A detailed study can be found in Refs.~\cite{Chien:2015xha,Brod:2018lbf}. With conservative values of matrix elements linking CP-odd quark-gluon operators to the neutron and Hg EDMs, we obtain $\Lambda_t \geq 0.7$ TeV, which is significantly weaker than the eEDM constraints, while no significant constraints can be set on $\Lambda_{b}$. For completeness we also give the EDM constraints for lighter quarks. Using conservative values for hadronic matrix elements there is no significant constraint for $\Lambda_s$ and $\Lambda_c$, while $\Lambda_d \geq 1$ TeV and $\Lambda_u \geq 0.5$ TeV \cite{Chien:2015xha,Brod:2018lbf}. Despite these weaker limits, we will see that EWBG is not efficient for CPV couplings involving light quarks.

The CPV fermion-Higgs couplings can also be probed at the LHC. At present, measurements of genuine CP-odd observables are not precise enough to set meaningful constraints. However, the CPV couplings modify also CP-even observables via contributions proportional to $ \tilde c_f^2 \sim 1/\Lambda_f^4$. For example, the CPV tau-Higgs coupling modifies the $h\rightarrow \tau\tau$ branching ratio signal strength
\begin{equation}\label{muhtautau}
\mu_{h\rightarrow \tau\tau}=\frac{\Gamma_{h\rightarrow \tau\tau}}{\Gamma^{SM}_{h\rightarrow \tau\tau}}\frac{\Gamma^{SM}_{h}}{\Gamma^{}_{h}} = 1 + \frac{v^4}{\Lambda_\tau^4}\left(\frac{1}{1-4 x_\tau}\right)\left(1-\frac{m_H m_\tau^2}{8\pi v^2}\frac{\left(1-x_\tau\right)^{3/2}}{\Gamma^{\rm SM}_{h}}\right)\,,
\end{equation}
where $\Gamma_h$ denotes the total Higgs width in presence of the CPV operators and $\Gamma_h^{\rm SM} \simeq 4.1$ MeV the predicted SM Higgs width. $\mu_{h\rightarrow \tau\tau}$ has been measured by ATLAS, ${\mu_{h\rightarrow \tau\tau} =1.09^{+0.18+0.27+0.16}_{-0.17-0.22.-0.11}}$, \cite{ATLAS:2018lur}. Adding the uncertainties in quadrature gives the constraint $\Lambda_\tau  \gtrsim 0.3$ TeV. Such limits are thus not very stringent. Similar analyses can be performed for other fermions, but in all cases the bounds on $\Lambda_f$ are well below $1$ TeV \cite{Chien:2015xha}.

\subsection{Transport equations}\label{sec:transport}
In this section we present the quantum Boltzman transport equations for a system with CP-violating sources for top and bottom quark as well as the tau lepton, generalizing the results in Refs.~\cite{Lee:2004we,Chung:2009qs}. From these equations we can compute the density of left-handed particles. This density sources the electroweak sphaleron transition that generates a net baryon number. We denote the net number density --- meaning the number density of particles minus antiparticles ---  of third-generation quarks  by $t= n_{t_R}, \, b = n_{b_R},\, q=n_{t_L} + n_{b_L}$,  the third-generation leptons  by $\nu= n_{\nu_R}, \, \tau = n_{\tau_R},\, l=n_{\nu_L} + n_{\tau_L}$, and simlarly for the lighter generations; the Higgs number density is given by $h= n_{H^0} + n_{H^+}$.

A careful analysis of the relevant timescales for the creation of the chiral asymmetry is given in section \ref{s:tau}. 
Gauge interactions and Higgs self-interactions are fast compared to the relevant time scales and are therefore assumed to be in thermal equilibrium, implying that the chemical potentials of the up and down components of $SU(2)_L$ doublets are equal. The same holds for the components of the Higgs doublet. First- and second-generation Yukawa interactions are slow and are therefore neglected; we justify this approximation in this work.  Consequently, the light leptons effectively decouple. The light quarks still participate in strong sphaleron interactions, which means their densities are related via
\be
q_1=q_2 = -2 u = -2d = -2s =-2c\,,
\label{ss}
\ee
and we only require one equation to describe them.  We choose the $u$-quark.  Here $q_i$ denotes the first- and second-generation left-handed doublet, and $u,d,s,c$ the right-handed quarks.
If we neglect the bottom Yukawa interactions, we have the further simplification $u=b$.

Weak sphaleron processes are also slow and baryogenesis can be modeled as a two-step process, where in the first step a chiral asymmetry is generated, which in  a second step is converted into a baryon asymmetry \cite{Carena:2002ss,Cline:2000nw}.  We argue  in \cref{s:tau} that this two-step approach even works for a lepton source scenario, where the relevant tau-Yukawa interaction rate can be small compared to the weak sphaleron rate. 

With the above considerations the full set of transport equations becomes
\begin{align}
\partial_\mu q^\mu &= 
+\Gamma_M^{(t)}\mu_M^{(t)}
+ \Gamma_M^{(b)}\mu_M^{(b)}
+ \Gamma_Y^{(t)} \mu_Y^{(t)}
+\Gamma_Y^{(b)} \mu_Y^{(b)}
 -2 \, \Gamma_{\rm ss} \, \mu_{\rm ss}
+\Gamma_{\rm QL}  \,\mu_{\rm QL} - S_t-S_b \, ,
\nn\\
\partial_\mu t^\mu  &=
-\Gamma_M^{(t)}\mu_M^{(t)} -  \Gamma_Y^{(t)} \mu_Y^{(t)}
+ \Gamma_{\rm ss} \, \mu_{\rm ss}-\Gamma_{\rm QL} \,\mu_{\rm QL} +S_t\,,
\nn \\
\partial_\mu b^\mu  &=
-\Gamma_M^{(b)}\mu_M^{(b)}- \Gamma_Y^{(b)} \mu_Y^{(b)}
+ \Gamma_{\rm ss} \, \mu_{\rm ss} +S_b\,,
\nn  \\[0.5cm]
  \partial_\mu l^\mu &= 
+ \Gamma_M^{(\tau)}\mu_M^{(\tau)}
+\Gamma_Y^{(\tau)} \mu_Y^{(\tau)}
 -\Gamma_{\rm QL} \, \mu_{\rm QL}                      -S_\tau \, ,      
\nn \\
\partial_\mu \nu^\mu  &=0,
\nn \\
\partial_\mu \tau^\mu  &=
-\Gamma_M^{(\tau)}\mu_M^{(\tau)}- \Gamma_Y^{(\tau)} \mu_Y^{(\tau)}
+\Gamma_{\rm QL}  \,\mu_{\rm QL} +S_\tau\,,
                         \nn \\[0.5cm]
 \partial_\mu h^\mu &= 
                      +\Gamma_Y^{(t)} \mu_Y^{(t)}-\Gamma_Y^{(b)}  \mu_Y^{(b)}\, +\Gamma_Y^{(c)} \mu_Y^{(c)}-\Gamma_Y^{(\tau)}  \mu_Y^{(\tau)}\,,
  \nn\\[0.5cm]
  \partial_\mu u^\mu &= + \Gamma_{\rm ss} \mu_{\rm ss}\,.                       
 \label{transport}
\end{align}
In principle, we should also add a transport equation for the new light degrees of freedom added to the scalar sector. Generically, these new degrees of freedom  equilibrate with the SM Higgs as significant interactions are required for a first-order phase transition, and the scalar degrees of freedom  can  be added to $h$ which now denotes a combined number density. 

The $S_f$ denote flavor-diagonal CPV sources for third-generation fermions. The relaxation rates $\Gamma_M^{(f)}$, Yukawa rates $\Gamma_Y^{(f)}$, and strong sphaleron rate $\Gamma_{\rm ss}$, redistribute and/or wash out the generated chiral asymmetry.  If the new degrees of freedom also couple to SM fermions additional interaction rates can appear in the above equations. We model one such possible term using $\Gamma_{\rm{QL}}$ that corresponds to a new top-tau interaction. This coupling will be motivated in section \ref{s:toplepcompare}. The lepton sector mirrors the quark sector in the above equations, with the important difference  that the leptons do not have strong sphaleron transitions, and the right-handed neutrino decouples. The values of the interaction rates for our benchmark model are given in \cref{table:rates} in \cref{A:BM}.

The chemical potentials (strictly speaking, these are rescaled chemical potentials, as we have factored out a factor $6/T^2$) corresponding to the interaction rates $\Gamma_M^{(f)}$,  $\Gamma_Y^{(f)}$, $\Gamma_{\rm ss}$ and $\Gamma_{\rm QL}$ are
\begin{align}
\mu_M^{(t)}&=\left( \frac{t}{k_t} - \frac{q}{k_q} \right) ,&
\mu_Y^{(t)}&=\left(\frac{t}{k_t}-\frac{q}{k_q}-\frac{h}{k_{h}} \right),
\nn \\
\mu_M^{(b)}&=\left( \frac{b}{k_b} - \frac{q}{k_q} \right) ,&
 \mu_Y^{(b)}&=\left(\frac{b}{k_b}-\frac{q}{k_q}+\frac{h}{k_{h}} \right),
                                                                          \nn \\
\mu_M^{(\tau)}&=\left( \frac{\tau}{k_\tau} - \frac{l}{k_l} \right) ,&
                                                                      \mu_Y^{(\tau)}&=\left(\frac{\tau}{k_\tau}-\frac{l}{k_l}+\frac{h}{k_{h}} \right), \nn \\                                                                   
\mu_{\rm ss}&= \sum_{i=1}^3 \( \frac{2q_i}{k_{q_i}} -\frac{u_i}{k_{u_i}} -\frac{d_i}{k_{d_i}}\),&
\mu_{\rm QL} &= \( \frac{l}{k_l} -\frac{\tau}{k_\tau} -\frac{q}{k_q} + \frac{t}{k_t} \).
  \label{mu}                                                                                    
\end{align}
The $k_i(m_i/T)$-functions relating the chemical potentials to the number densities are defined via $n_i=T^2\mu_i k_i/6 + {\mathcal O}(\mu_i^3)$.

In the diffusion and planar-wall approximation for the bubble profile, we can write the left-hand-side of the transport equations as \cite{Cohen:1994ss}
\be
\partial_\mu n_f^\mu = v_w n_f' - D_f n_f''\,,
\ee
with $v_w$ the bubble-wall speed and $D_f$ the diffusion coefficient (listed in \cref{A:BM}).
Since the left- and right-handed quarks have approximately equal diffusion constants, baryon number is locally conserved on the time scale of the transport equations such that
\be
	t+ b + q + c + s + q_2 + u + d + q_1 =	t+ b + q  =0\,.\label{localbc}
\ee
In the second expression we used that the light quarks are only produced via strong sphalerons \cref{ss}.  Local baryon number conservation can be used to eliminate the transport equation for the bottom quark. The set of transport equations can be further simplified by neglecting the slower rates, but which rates can be neglected depends on the chosen source term.
If bottom Yukawa interactions are neglected we have $b = u$, and the number densities of all quarks directly follow from $(q,t)$. If bottom Yukawa interactions are included we additionally have to solve the $u$-equation.  

In contrast, lepton number is only conserved globally \cite{Chung:2009cb}. Right-handed leptons diffuse more easily than left-handed leptons since right-handed leptons do not interact through $SU_L(2)$-interactions, and therefore $D_l \neq D_\tau$. As we will see, for our set-up it is a reasonably good approximation to neglect this difference and assume local lepton number conservation as well.

The complete expressions for the interaction rates, source terms, masses, and other constants entering the equations can be found in Ref.~\cite{deVries:2017ncy}.  For the Yukawa rates we used the expressions in Ref.~\cite{Joyce:1994zn}. The rate $\Gamma_{\rm QL}$ is discussed in \cref{sec:GammaQL}.
We have solved the transport equations numerically, with the boundary condition that the number densities vanish far away from the bubble wall. Details are given in \cref{A:numerics}. As an extra check we also solved the equations using the semi-analytical method of Ref.~\cite{White:2015bva}. This method approximates all rates by a step function, and ignores the variation over the bubble-wall width. In addition, since the source peaks in the broken phase, it is set to zero in the symmetric phase.  Both approximations are reasonable, and for generic input parameters we find that the numerical and semi-analytical results only differ by ${\mathcal O}(10\%)$.  Unless otherwise stated, the results presented in the upcoming sections are those of the numerical calculation.

\subsection{Electroweak sphaleron transitions}
\label{s:EWS}
The electroweak sphalerons convert the chiral asymmetry into a baryon asymmetry. The corresponding rate is slower than all  other relevant interaction rates and thus decouples from the transport equations. The only exception is the lepton Yukawa rate and we discuss this case in \cref{s:tau}.
The  density of left-handed fermions that sources the electroweak sphaleron transitions is given by
$	n_L = \sum_i (q_i + l_i)\,,
$
and is determined by solving the transport equations in \cref{transport}. The baryon asymmetry becomes (see Appendix~\ref{s:onestep} for a derivation)
\be
Y_B = \frac{n_b}{s} = 
-\frac{3  \Gamma_{\rm ws}}{2 s D_q \alpha_+} \int_{-\infty}^0 \dd z \, n_L \e^{-\alpha_- z}\,,
\label{YB}
\ee
where
\be
\alpha_\pm = \frac{v_w \pm \sqrt{4 D_q \Gamma_{\rm ws} {\mathcal R} +v_w^2}}{2D_q}\,.
\label{alpha}
\ee
Here $s = 2 \pi^2/(45) g_{*S} T^3$ is the entropy density, $g_{*S} = 106.75$ the entropy degrees of freedom at the electroweak scale, $D_q \simeq 6/T$ the quark diffusion constant, ${\cal R} = 15/4$ the SM relaxation term, and $\Gamma_{\rm ws} = 6 \kappa \alpha_w^5 T$ the electroweak sphaleron rate with $\kappa \sim 20$ and $\alpha_w = g^2/(4\pi)$ \cite{Bodeker:1999gx,Moore:1999fs,Moore:2000mx}.  
In the limit $4 D_q  \Gamma_{\rm ws} {\mathcal R} \ll v_w^2$, the result reduces to a more familiar form
\be
\label{nb_sol}
Y_B = \frac{n_b}{s}= - \frac{3 \Gamma_{\rm ws}}{2 v_{w}s} \, \int_{-\infty}^{0}  \ 
n_L (z) 
\,  e^{z \, {\cal R} \Gamma_{\rm ws}/v_w}\, .
\ee
This approximation works well for $v_w \geq 0.02$.
We integrate the asymmetry over the broken phase ranging from $z= -\infty$ to the center of the bubble wall at $z=0$ where $\phi_b = v_N/2$.  Other integration regions can be chosen, for instance $-\infty < z <-L_w$ \cite{Lee:2004we, Chung:2009cb}.  This gives a percent-level difference for the asymmetry generated by a lepton source, but can give an $\mathcal O (1)$ difference for the top source.  It would be optimal to integrate over the full region using the field-dependent $\Gamma_{\rm ws}(\phi_b)$, but this requires a better understanding of the electroweak sphaleron rate.

\subsection{Efficiency of quark/lepton source}
\label{s:toplepcompare}
A priori, one would expect the top source to give the largest baryon asymmetry. Since the CPV source term of \cref{Sf2} is proportional to the Yukawa coupling squared, it is maximal for the top quark. Even if we had not assumed the dimension-six couplings to be proportional to $y_f$, the source would have a linear dependence on the Yukawa coupling. 
A first reason why the top-source scenario might nevertheless not be the most efficient mechanism to generate the BAU is also immediately apparent from \cref{Sf2}. The CPV source is suppressed by the square of the scale $\Lambda_f$.
In section \ref{sec:EDM} we showed that the experimental bounds on $\Lambda_f$ are rather strong for the top quark: $\Lambda_t  \gtrsim 7.1 \, \text{TeV}$. For the bottom quark and the tau lepton the bounds are much less severe: $\Lambda_b \gtrsim 0.5 \, \text{TeV}$ and $\Lambda_\tau \gtrsim 0.3 \, \text{TeV}$ respectively.

A second reason why leptons could be more efficient than quarks in generating the BAU is their larger diffusion coefficient. Since leptons only interact via the electroweak force they can diffuse further into the symmetric phase than the strongly interacting quarks \cite{Joyce:1994bi}. This enhances the baryon asymmetry, as the electroweak sphalerons have more time to convert the lepton asymmetry into a baryon asymmetry before the bubble wall passes.

Finally, washout effects are also maximal for top quarks. The interactions mediated by the CP conserving part of the mass matrix relax the chiral asymmetry; the relaxation rate for this process is proportional to $\Gamma^{(f)}_M \propto |\bar m_f|^2 = y_f^2 \phi_b^2 + {\mathcal O}(\Lambda_f^{-4})$, and thus largest for the top quark. The Yukawa-type interactions are kinematically forbidden in the plasma, but phase space opens up if an additional gluon or weak boson is radiated; the rate for this process is likewise maximized for the top quark $\Gamma_Y^{(f)} \propto  y_f^2$.  

More importantly, the strong sphaleron transitions are approximately in equilibrium in the symmetric phase except for regions close to the bubble wall \cite{Giudice:1993bb, Tulin:2011wi,Huet:1995sh} --- we will estimate the size of this region in \cref{s:top} ---  and they very effectively wash out the chiral asymmetry in the quark sector, but leave leptons untouched. Indeed, if the strong sphaleron interactions are in thermal equilibrium the corresponding chemical potential \cref{mu} vanishes $\mu^{ss} \simeq 0$.  Neglecting the Yukawa interactions of the first- and second-generation quarks, and using baryon number conservation gives the relation
\be
 -4 u a_1+q a_2 + b a_3 \simeq 0\,,
\label{ss_eq}
\ee
where
\be
a_1 =\frac12\( \frac{4}{k_{q_1}} +\frac1{k_u} +\frac1{k_d}\)\,, \quad
a_2 = \( \frac{2}{k_{q}} +\frac1{k_t} \)\,, \quad
a_3 = \( \frac{1}{k_{t}} -\frac1{k_b} \)\,.
\ee
The chiral asymmetry in quarks becomes%
\be
n_L^{(q)} \equiv \sum_{i=1}^3 q_i = q -4 u = q \(1-\frac{a_2}{a_1} \) - b\(\frac{a_3}{a_1}\)\,,
\label{nlq_low}
\ee
where in the second and third steps we used \cref{ss} and \cref{ss_eq}, respectively.  At zero temperature $(a_1,a_2,a_3) \propto(1,1,0)$ and the chiral asymmetry in quarks vanishes. At finite temperature there are corrections.  For our benchmark point we find $n_L^{(q)} \simeq -0.01 q-0.01 b$, and thus the chiral asymmetry in quarks is suppressed by roughly two orders of magnitude with respect to the individual quark densities. 

The above reasons ensure that the tau-source scenario is more effective at producing the baryon asymmetry than the top-source scenario. That is, despite the Yukawa suppression of the tau source we will obtain similar-sized total baryon asymmetries for $\Lambda_t \simeq \Lambda_\tau$.

The top Yukawa and especially the strong sphaleron interactions are very effective in washing out the chiral asymmetry in quarks.  Hence, if some of the chiral asymmetry can be transferred to the leptons, which will escape the washout, the baryon asymmetry is increased.  Leptons produced via the relatively small tau Yukawa interactions give the dominant contribution to the baryon asymmetry in the top-source scenario leading to a larger asymmetry up to an order of magnitude (see fig.~\ref{fig:top_vw}).
The authors of Refs.~\cite{Chung:2008aya, Chung:2009cb, Joyce:1994bi} already pointed out that the contribution of tau leptons can be significant  in the context of a two Higgs doublet model and the MSSM where the value of $y_\tau$ can be boosted by a large $\tan \beta$, but we stress that this Yukawa enhancement is not required for leptons to be very relevant.  This relevance can be even enhanced in models with additional chiral-symmetry-breaking lepton-quark interactions, for instance via the exchange of new scalars. We discuss this in the next section.

\subsection{Additional chiral-symmetry-breaking quark-lepton interactions}
\label{sec:GammaQL}
The importance of the leptons in the top-source case becomes even more pronounced in models with additional chiral-symmetry-breaking quark-lepton interactions. In our set-up we have added a dimension-six  tau-top interaction to study this effect, but the qualitative results are insensitive to the exact implementation. The dimension-six operator is given by
\be
{\mathcal L}_{\rm QL} =\frac{1}{\Lambda_{\rm QL} ^2} \bar \tau_L \tau_R \bar t_R t_L + {\rm h.c.} \, .
\label{L_QL}
\ee
Using the methods of Ref.~\cite{Joyce:1994zn} we have calculated the rate associated to this interaction and obtain $\Gamma_{\rm QL}  = \kappa_{\rm QL}{T^5}/{\Lambda_{\rm QL} ^4}$ with $\kappa_{\rm QL}$ a factor of $\mathcal O(1)$. This interaction becomes important if it exceeds the SM tau Yukawa rate, which, for our benchmark point, is the case for $\Lambda_{\rm QL}  \lesssim 3 \,$TeV.

For definiteness, we focus here on a new top-tau coupling but stress that a coupling to the muon or electron has a similar effect. If the CPV source is located in the quark sector,
 the baryon asymmetry can also be boosted by including a new coupling of the top to one of the lighter quarks (e.g. a top-charm coupling). In such a scenario the washout by strong sphalerons becomes less effective.  Indeed, \cref{nlq_low} above was derived under the assumption that the light quarks are only produced via strong sphalerons and that their Yukawa interactions are negligible, such that \cref{ss} holds. If, say, the new top-charm interaction is stronger than the strong sphaleron interactions, this is no longer the case, and the washout by strong sphalerons no longer implies a washout of the chiral asymmetry.

We describe the top-tau transfer here with an effective interaction, but larger interaction rates are possible if the coupling is via exchange of a light degree of freedom, for example the new scalar particle added to get a first-order phase transition \cite{Chiang:2016vgf, Guo:2016ixx}. In such cases the scattering can be resonantly enhanced.  To properly describe this requires the inclusion of the transport equation for the light particle with the corresponding interaction rates derived from renormalizable interactions. We expect that our effective interaction gives qualitatively similar results, even in the regime where $\Lambda_{\rm QL}$ is fairly small, and we therefore treat $\Lambda_{\rm QL}$ as a phenomenological parameter. We will show in \cref{s:extra} that such interactions, even of modest strength, can drastically increase or reduce the baryon asymmetry depending on the nature of the CPV source.

\section{Baryogenesis with a tau-lepton source}
\label{s:tau}

We start with a discussion of baryogenesis from a CPV tau source, and assume there is no new lepton-quark coupling (that is, we set $\Lambda_{\rm QL} \to \infty$). The value of the tau Yukawa coupling at the electroweak scale is small, $y_\tau \simeq 0.01$, with respect to the top Yukawa coupling, $y_t \simeq 1$.  For reasons discussed in section \ref{s:toplepcompare}, the tau could nevertheless be an interesting source of the baryon asymmetry. 

In the present scenario, the lepton sector essentially decouples from the quark sector and the baryon asymmetry can be computed analytically to good accuracy. We start with discussing the analytical approximation and later compare it to the full numerical solution that does include effects of the quark sector. The analytical solution clarifies the dependence of the baryon asymmetry on parameters associated to the bubble wall and the leptonic Yukawa couplings. This relatively simple set-up provides insight into more complicated scenarios where analytical solutions are not possible.

\subsection{Analytical approximation}
 
The transport equations in \cref{transport} contain separate equations for the third-generation left-handed doublet, $l^\mu$, and the right-handed singlet, $\tau$. Because the left- and right-handed tau leptons diffuse at different rates:  $D_l = 100/T$ and $D_\tau =380/T$ \cite{Chung:2009cb}, in principle we have to treat these number densities separately.  However, to a reasonable approximation we can ignore this difference and set $D_l = D_\tau = 100/T$, which implies $l = -\tau$. The lepton transport equation simply becomes
\be
- D_l l''+ v_w l' + \bar \Gamma l -\Gamma_Y^{(l) }\frac{h}{k_h} = -S_\tau\,,
\qquad {\rm with} \qquad \bar \Gamma = \( \Gamma_M^{(l)}  
+\Gamma_Y^{(l)} \)\(\frac{1}{k_\tau}+\frac{1}{k_l}\)  .
\label{eql}
\ee
CPV resides in the lepton sector and any non-zero density of Higgs particles can only be generated via interactions with leptons. Since the lepton Yukawa rate is relatively slow the Higgs density remains small and we can approximate $h \simeq 0$. In this limit, the leptons decouple completely and to find the chiral asymmetry we only have to solve \cref{eql}. This can be done using the semi-analytical solution mentioned at the end of section \ref{sec:transport}, which gives%

\begin{equation}
l = -\frac{2\bar S}{(\sqrt{4 D_l \bar \Gamma_B +v_w^2}+\sqrt{4 D_l \bar \Gamma_S +v_w^2})}
\e^{ z/L_S}\,, \quad {\rm for} \;\; z <0\,,
\label{sol_l} 
\end{equation}
with $\bar \Gamma_{\{S,B\}}$ the rescaled interaction rates \cref{eql} in the symmetric and broken phase respectively, and
\begin{align}
  L_i &= \frac {2D_l}{\(v_w + \sqrt{4 D_l \bar \Gamma_i + v_w^2}\)}\,,\nn\\ 
    \bar S &= \int_0^\infty \e^{- y/L_B} S_\tau(y) \dd y   =  \frac{15s_\tau}{128\pi^2} \frac{y_\tau^2}{\Lambda_\tau^2} v_w v_N^4 J_\tau(T) \left[1 -\frac{4(7+6 \ln2)}{45} \frac{L_{w}}{L_B}  + {\mathcal O}\(\frac{L_w^2}{L_B^2}\)\right]\, ,
  \label{barS}
      \end{align}
and $i=\{S,B\}$.  To get the second expression for $\bar S$, we used eqs.~\eqref{tanh} and \eqref{Sf2}, and expanded in $L_w/L_B$. For a tau source, the leading order term dominates by far and we can set $L_w /L_B \simeq 0$.
      
Integrating the chiral asymmetry, $n_L = l$, as in \cref{nb_sol} gives the baryon asymmetry
\begin{equation}\label{Ybtau}
  Y_B
  = \frac{3 \Gamma_{\rm ws}}{ 2 v_w s} \frac{2\bar S}{(\sqrt{4 D_l \bar \Gamma_B +v_w^2}+\sqrt{4 D_l \bar \Gamma_S +v_w^2})}\, \frac{L_{\rm ws} L_S}{L_S + \frac{1}{2}L_{\rm ws} \left(1+\sqrt{1+\frac{4 D_q}{L_{\rm ws} v_w}}\right)}\, ,
\end{equation}
with $L_{\rm ws} = v_w/({\mathcal R} \Gamma_{\rm ws})$.   Noting that $J_\tau <0$, we have to pick $s_\tau =-1$ for the sign of the CPV coupling, to get the right sign for the baryon asymmetry.
The last factor in \cref{Ybtau} arises from 
\be
I_Y \equiv \int \dd z \, \e^{z \left(L_S^{-1} - \alpha_- \right)} = \frac{L_{\rm ws} L_S}{L_S + \frac{1}{2}L_{\rm ws} \left(1+\sqrt{1+\frac{4 D_q}{L_{\rm ws} v_w}}\right)}\, .
\label{IY}
\ee 

 \subsection{Comparison of approximations}
 
We have calculated the total baryon asymmetry using different approximations:
\begin{enumerate}
\item The analytical solution in \cref{Ybtau} of the purely leptonic transport equation, \cref{eql},  where the Higgs and quark sectors are neglected. We call this solution $A(l)$. 
\item A numerical solution of \cref{eql} called $N(l)$. 
\item A numerical solutions that includes the Higgs and third-generation quarks called $N(l,h,q,t)$. 
\item Finally, we investigate the different diffusion constants of left- and right-handed leptons. That is,  we allow $D_l \neq D_\tau$ in $N(l,\tau)$. For numerical reasons we study this effect in the limit of a decoupled quark sector.
\end{enumerate}
The entries in brackets in $A(...)$ and $N(...)$  denote the number densities for which we solve the transport equation explicitly in the respective approximation. In scenarios where the quark sector is considered, we apply local baryon conservation, \cref{localbc}, and the relation in \cref{ss} to account for the number densities of the bottom and lighter quarks.

The resulting baryon asymmetries  in the various approximations for our benchmark point with $\Lambda_\tau = 1\,$TeV and bubble-wall velocity $v_w =0.05$ are given in  \cref{table:tau_source}. Comparing N$(l,\tau)$ to N$(l)$ shows that local lepton conservation is a reasonable approximation. More importantly, the analytical solution differs only $\sim 10\%$ from the numerical solutions  withand without inclusion of the Higgs and quark sectors, that is from N$(l,h,q,t)$ and  N$(l)$ respectively. The tau Yukawa coupling is sufficiently small that decoupling the Higgs and quark sectors is an excellent approximation. 

In the left panel of  \cref{fig:tau_source} the various approximations are compared while varying the value of the tau Yukawa coupling. The decoupling approximation works well up to $y_\tau \lesssim 0.2$ while for larger values significant Higgs and quark densities are generated. The right panel of \cref{fig:tau_source} shows that the analytical approximation holds over a large range of bubble-wall velocities, such that the conclusions are not restricted to just our benchmark values.

\begin{table}[t]
\renewcommand{\arraystretch}{1.2}
	\centering
	\begin{tabular}{ | l || c | c |c|c| }
	\hline 
          Approximation & A$(l)$  &N$(l)$  & N$(l,h,q,t)$  &N$(l,\tau)$ \\
          \hline
  $Y_B$ & $6.8\times 10^{-11}$ & $7.3 \times 10^{-11}$ & $7.3 \times 10^{-11}$ & $8.2 \times 10^{-11}$  \\\hline 
    \end{tabular}
    \caption{
      Baryon asymmetry $Y_B$ for a tau CPV source obtained using various approximations as detailed in the main text. We used the benchmark values for the bubble-wall profile given in \cref{sec:FOPS} and $\Lambda_\tau = 1$ TeV. }
    \label{table:tau_source}
\end{table}

\subsection{Parameter dependence}\label{sec:pardep}
Having found that the analytical solution provides an excellent approximation, we can use it to understand how the baryon asymmetry depends on various parameters, such as the Yukawa coupling and the parameters associated to the bubble wall. It is useful to rewrite the solution for $Y_B$ in \cref{Ybtau}  in terms of various length scales
\begin{equation}\label{Ybtau2}
  Y_B
  = \frac{3}{ s {\mathcal R}} \frac{\bar S}{v_w L_{\rm ws} } \frac{1}{\sqrt{1+4\frac{L_{\rm diff}}{L^{S}_{\rm int}}}+\sqrt{1+4\frac{L_{\rm diff}}{L^{B}_{\rm int}}}}\,\,\frac{L_{\rm ws} L_S}{L_S + \frac{1}{2}L_{\rm ws} \left(1+\sqrt{1+4\frac{ L^q_{\rm diff}}{L_{\rm ws}}}\right)}\, ,
\end{equation}
which are defined in \cref{table:lengthscales}.
\begin{table}[t]
\renewcommand{\arraystretch}{1.15}
	\centering
	\begin{tabular}{ | l |l|r| }
	\hline 
        Length scale& Description  & Benchmark value \\
          \hline \hline
          $L_S=\frac{2L_{\rm diff}}{1+\sqrt{1+{4L_{\rm diff}}/{L_{\rm int}^S}}}$ & migration length in symmetric phase & 21.3\\ \hline
          $L_B=\frac{2L_{\rm diff}}{1+\sqrt{1+{4L_{\rm diff}}/{L_{\rm int}^B}}}$ & migration length in broken phase &9.8 \\ \hline
          $L^S_{\rm int}=v_w/\bar \Gamma_S$ & interaction length in symmetric phase & 313.3\\ \hline
            $L^B_{\rm int}=v_w/\bar \Gamma_B$ &interaction length in broken phase & 7.0\\ \hline
          $L_{\rm diff}=D_l/v_w$ & lepton diffusion length & 22.7\\ \hline 
           $L^q_{\rm diff}=D_q/v_w$ & quark diffusion length & 1.4\\ \hline         
          $L_{\rm ws}=v_w/({\mathcal R} \Gamma_{\rm ws})$ & weak sphaleron length & 28.5\\ \hline
           $L_{w}$ & bubble-wall width & 0.1\\ \hline   
    \end{tabular}
    \caption{
      Length scales (in units of GeV$^{-1}$) for a tau CPV source and their values for the benchmark values for the bubble-wall profile given in \cref{sec:FOPS} and $\Lambda_\tau = 1$ TeV.
    }
    \label{table:lengthscales}
\end{table}
$L_S$ and $L_B$ determine how far the asymmetry migrates into, respectively, the symmetric and broken phase. On larger length scales the asymmetry is exponentially suppressed by the $\e^{ -|z|/L_i}$-factor in \cref{sol_l,barS}.  $L_{\rm diff} =D_l/ v_w$ is the diffusion length, which determines how far the asymmetry can diffuse into the symmetric phase before it is overtaken by the expanding bubble.  

The interaction lengths in the symmetric and broken phase are denoted by $L^{i}_{\rm int} = v_w/\bar \Gamma_{i}$ with $i=S,B$, respectively.   If interactions are slow, $L_{\rm diff} \ll L^{i}_{\rm int} $, they can be neglected, and the migration scale is determined by  diffusion  $L_i \simeq L_{\rm diff}$. In the opposite limit of fast interactions, the symmetry is washed out over scales larger than $L_i \simeq \sqrt{L_{\rm diff} L^i_{\rm int}} =\sqrt{D_l/\bar \Gamma_i} $. In the broken phase,  $L_B$ has to be compared with the spatial extend of the source, which is set by the bubble-wall width $L_w$.  For $L_w \ll L_B  $ the source is not diluted by diffusion nor interactions and we can approximate $\bar S$ by the first term in \cref{barS}. 

The conversion of the chiral number into a net baryon number through \cref{nb_sol} introduces additional length scales: the baryon/quark diffusion length scale $L_{\rm diff}^q =D_q/v_w$, and the scale on which the weak sphalerons act $L_{\rm ws}=v_w/({\mathcal R} \Gamma_{\rm ws})$.  In the limit $L_S \ll L_{\rm ws} $ the migration length, i.e., how far the chiral asymmetry migrates into the symmetric phase, determines the extend of the region in front of the bubble wall that contributes to the baryon asymmetry and the integral \cref{IY} becomes $ I_Y \simeq L_S $. In the opposite limit,
$L_{\rm ws} \ll L_S$, it seems necessary to include the weak sphaleron interactions directly in the transport equations, as there is no hierarchy between the sphaleron and Yukawa interactions, invalidating the two step procedure. However, although both processes may be important for the final asymmetry (depending on the bubble-wall velocity) they are still very far from equilibrium, and back reaction effects are small.  We checked numerically that the two-step procedure used to derive \cref{Ybtau} is a good approximation to the full coupled equations, see \cref{s:onestep} for more details.  We then find that in this regime the integral is cut off by $L_{\rm ws}$, the scale on which the weak sphalerons act, and $I_Y \simeq L_{\rm ws} (1+\sqrt{r})^{-1}$ in the parameter regime of interest, with
\be
r = \frac{L_{\rm ws} L_{\rm diff}^q}{L_S^2}  \; \stackrel{v_w \to 0}{=} \; \frac{D_q}{D_l} \frac{\bar \Gamma_S}{\mathcal{R} \Gamma_{\rm ws}}\,,
  \label{r}
\ee
where the right-hand expression is valid in the small-velocity limit where $r$ is maximal.  For the tau Yukawa interactions $r \ll 1$ and we can neglect the baryon diffusion effects, and the approximation made in \cref{nb_sol} holds. This is no longer the case for the top and bottom-source scenarios discussed in \cref{sec:quarks}, as can be anticipated from the above estimate. Replacing the tau Yukawa interaction with the top Yukawa or strong sphaleron interaction, and setting $D_l \to D_q$ gives $r \gg 1$.

The various length scales are listed in table \ref{table:lengthscales}, along with their value for the benchmark point. The length scales depend on bubble-wall parameters such as $v_N$ and $v_w$ that we can vary depending on the electroweak phase transition, and on SM parameters such as the tau Yukawa coupling, $y_\tau$. We also consider variations in $y_\tau$ to see what would happen for source terms involving lighter SM fermions and to get some (limited) insight about what happens for heavier quarks, as further discussed in the next section. 
For all parameter choices, we have $L_{\rm int}^B \ll L_{\rm int}^S$ as $\Gamma_M^{(l)}$ greatly exceeds the Yukawa interaction rate in both the symmetric and broken phase.  The solution for the baryon asymmetry can then be  divided into three\footnote{In principle there exists a fourth region $L_{\rm int}^S \ll L_{\rm diff}$ and $L_S = \sqrt{L_{\rm int}^S L_{\rm diff}} \ll L_{\rm ws}$.  This requires $\mathcal O(1)$ Yukawa couplings, for which our analytical approximation \cref{sol_l} breaks down. As this is not the physical region of interest, we do not discuss this possibility further.}
 different regions, depending on the chosen parameters and the resulting sizes of the relevant length scales:
\begin{enumerate}
\item[\textbf{a}\,\,\,]{$ L_{\rm diff}  < L_{\rm int}^B,L_{\rm int}^S,L_{\rm ws}\,$ which corresponds to small Yukawa couplings and a large bubble-wall velocity. } Interactions are slow and can be neglected, and the chiral asymmetry diffuses without washout $L_i \simeq L_{\rm diff}$. Since the diffusion length is much larger than the source width $L_{\rm diff} \gg L_w$ there is no dilution of the source and the expansion 
  in \cref{barS} is valid.
In this regime, the baryon asymmetry becomes
\begin{equation}
  Y_B \simeq \frac{3 }{2 s {\mathcal R}} \frac{\bar S}{v_w} \frac{L_{\rm diff}}{L_{\rm ws} } \propto \frac{y_\tau^2}{v_w^2}\frac{1}{\Lambda_\tau^2}\,.
\end{equation}
The asymmetry decreases for large bubble-wall velocity, which can be understood as the faster the bubble moves, the less time there is for the chiral asymmetry to diffuse into the symmetric phase and to be converted into a baryon asymmetry. The asymmetry decreases quadratically with a smaller Yukawa coupling, which originates from the scaling of the source. In practice, the tau Yukawa coupling is slightly too large for the above scaling to hold for the benchmark velocity $v_w \simeq 0.05$.

\item[\textbf{b}\,\,\,]{
    $L_{\rm int}^B < L_{\rm diff}<L_{\rm int}^S,L_{\rm ws}\,$.} Interactions are now important in the broken phase and $L_B\simeq \sqrt{L_{\rm int}^B L_{\rm diff}}$. The dilution of the source in $\bar S$ is still a negligible effect, as $L_B \simeq L_w$ only holds for $\mathcal O(1)$ Yukawa couplings.
 The interactions in the broken phase, however, change the baryon asymmetry into
\begin{equation}
  Y_B \simeq \frac{3 }{2 s {\mathcal R}} \frac{\bar S}{v_w} \frac{\sqrt{L_{\rm diff}L_{\rm int}^B }}{L_{\rm ws}}  \propto \frac{y_\tau}{v_w}\frac{1}{\Lambda_\tau^2}\,.
  \label{b1}
\end{equation}
$Y_B$ thus scales linearly with the Yukawa coupling and inversely with the bubble velocity. Here we see for the first time a violation of the naive $y_f^2$ scaling of the asymmetry.

\item[\textbf{c}\,\,\,]{$L_{\rm int}^B < L_{\rm ws} < L_{\rm diff},L_{\rm int}^S$ which corresponds to small velocities (and $y_\tau \lesssim 0.2$ such that the analytical approximation is valid).}
The baryon asymmetry 
\begin{equation}
  Y_B \simeq \frac{3 }{2 s {\mathcal R}} \frac{\bar S}{v_w} \sqrt{\frac{L_{\rm int}^B}{ L_{\rm diff}} }  \propto \frac{y_\tau v_w}{\Lambda_\tau^2}\,,
  \label{b2}
\end{equation}
now scales linearly with both $y_\tau$ and $v_w$.   The asymmetry decreases for small velocity,  as in this limit the evolution approaches thermal equilibrium.  We neglect the $L^S_{\rm int}$ dependence in the denominator in \cref{Ybtau} as it is subdominant to the $L_{\rm int}^B$ term.  

\end{enumerate}

Comparing \cref{b1,b2} it follows that the asymmetry is maximized as a function of the velocity at the boundary of the two regimes, that is, for $L_{\rm ws} \sim L_{\rm diff}$.  For the SM tau Yukawa coupling, region \textbf{a} corresponds to $v_w \gtrsim 0.1$, region \textbf{b} to $ 0.04 \lesssim v_w \lesssim 0.1$, and region \textbf{c} to $v_w \lesssim 0.04$. 

\begin{figure}
	\centering
	\includegraphics[width = 0.46 \linewidth]{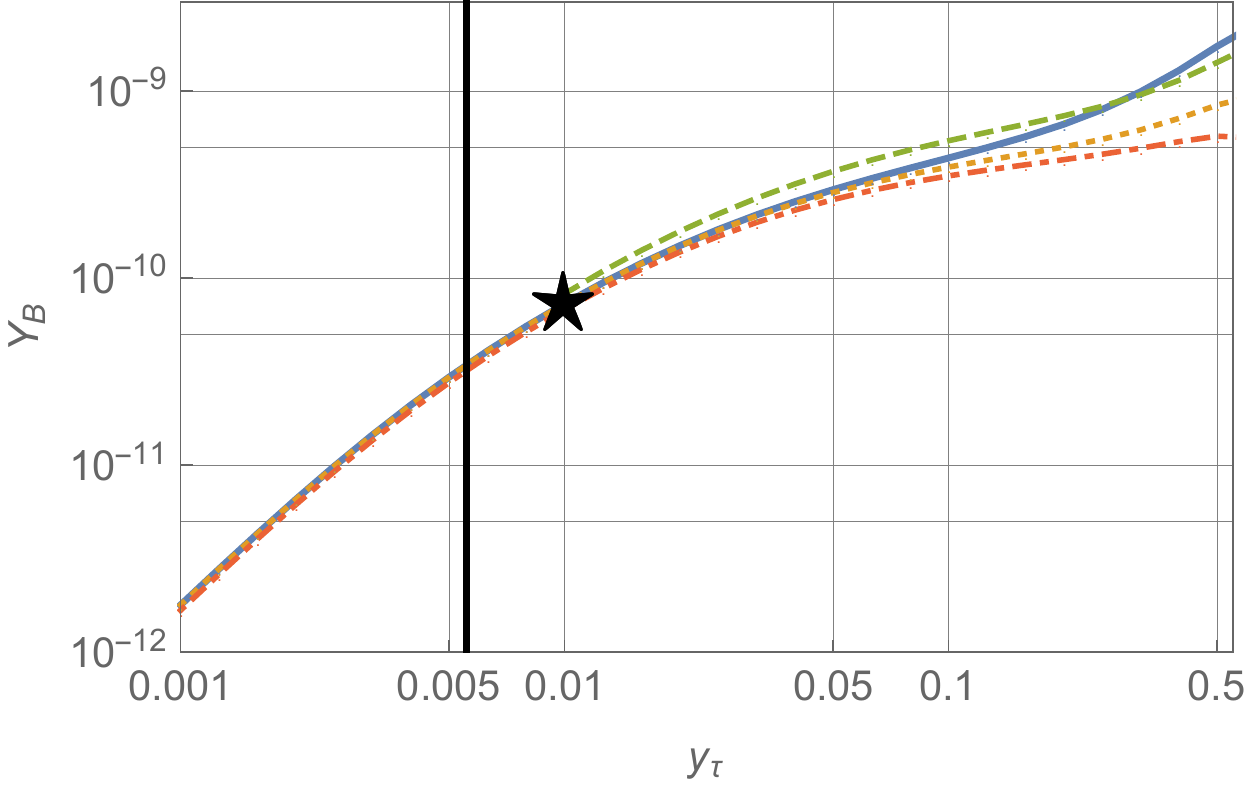}
		\includegraphics[width = 0.48 \linewidth]{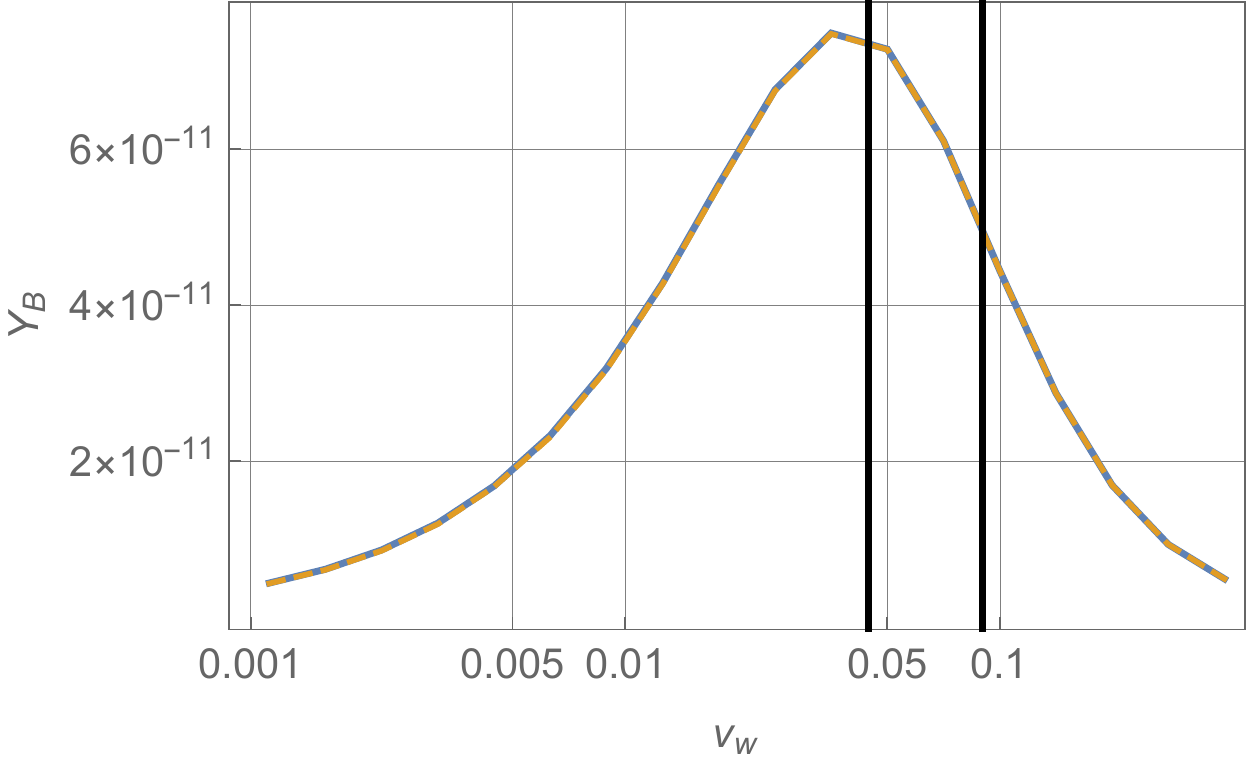}
	\caption{ Left: $Y_B$ as a function of the Yukawa coupling $y_\tau$ for the same four approximations as in \Cref{table:tau_source} (solid blue: N$(l,h,q,t)$, dotted orange: N$(l)$, dashed green N$(l,\tau)$, dotdashed red: A$(l)$). The region left (right) of the vertical black line corresponds to  region  \textbf{a} (\textbf{b}). The star denotes the SM value of $y_\tau$. \newline Right:  $Y_B$ as a function of the bubble-wall velocity $v_w$ (solid blue: $N(l,h,q,t)$, dashed orange: N$(l)$). The left, middle and right regions corresponds to regions \textbf{c},  \textbf{b}, and  \textbf{a}, respectively.}
	\label{fig:tau_source}
\end{figure}

The scaling of the asymmetry with bubble-wall velocity and Yukawa coupling are illustrated in \cref{fig:tau_source}.  The left panel shows $Y_B$ as a function of the Yukawa coupling $y_\tau$. The vertical thick line gives the boundary between region \textbf{a} and \textbf{b};   region \textbf{c} does not occur for our benchmark velocity.   In region \textbf{a}, on the left of the vertical thick line, the baryon asymmetry scales quadratically with the small Yukawa coupling, while in region \textbf{b}, on the right, the scaling is linear.  The SM Yukawa value $y_\tau =0.01$ falls in region \textbf{b}, and the baryon asymmetry for this value is indicated by a star (and corresponds to our benchmark number in \cref{table:tau_source}). The most relevant observation is that for  $y_\tau \geq 0.005$, $Y_B$ only grows linear with $y_\tau$ instead of the naively expected quadratic scaling. This confirms that CPV sources for fermions lighter than top quarks are less disadvantageous than might be expected.

The right panel shows $Y_B$ as a function of the bubble-wall velocity $v_w$. Now the black lines divide, from right to left, regions \textbf{a}, \textbf{b}, and \textbf{c} that scale respectively as $v_w^{-2}$, $v_w^{-1}$ and $v_w$.  The boundary between \textbf{b} an \textbf{c} correspond to the velocities for which $L_{\rm ws} \sim L_{\rm diff}$, close to which the baryon asymmetry is maximized. Our benchmark point lies in scenario \textbf{b}, very close to the optimum value.

We end this section with a short discussion of how the baryon asymmetry depends on the dynamics of the first-order phase transition that we have parameterized by the bubble-wall profile \cref{tanh}.  The bubble-wall width only enters the equations via the source.  In the limit that $L_w /L_B \ll 1$, as appropriate for a tau source, the dependence on this parameter cancels.  A larger bubble width gives a wider source but with a smaller amplitude, such that the total integrated source $\bar S$ remains the same.  

The source scales as $\bar S \propto v_N^4$ with $v_N$ the Higgs vev during nucleation; this power comes about because of the structure of our dimension-six CPV operator which involves three Higgs fields\footnote{In two-Higgs doublet models CPV can originate from dim-4 operators, in which case the corresponding source would only scale quadratically with the vev.}.   The relaxation rate $\Gamma_M^{(f)} \propto v_N$ depends on the vev as well.  It thus follows that in  region \textbf{a}, the baryon asymmetry scales as $Y_B \propto v_N^4$ and in  region \textbf{b}  as $Y_B \propto v_N^3$. For the SM tau Yukawa coupling we are in region \textbf{b}, but the velocity dependence of $L_B$ cannot be fully neglected and numerically we find a scaling $Y_B \propto v_N^{3.2}$.  Although the dependence on the vev is strong, we do not expect $v_N$ to vary too much between different BSM models.  

The dependence on the nucleation temperature $T_N$ is harder to estimate, as many parameters, such as the thermal masses and thermal width, depend on the temperature. All interaction rates scale $\Gamma_i \propto T$, except for $\Gamma_M \propto 1/T$; further $D_i \propto 1/T$ and $S \propto T$. With this scaling we find that $Y_B \propto 1/T^2_N$ in regime \textbf{a} and $Y_B \propto 1/T_N$ in regime \textbf{b}. We note that as $T_N$ increases the relative importance of $L_{\rm int}^B$ decreases, and we move from regime \textbf{b} to \textbf{a}.

\subsection{Producing the universal baryon asymmetry with a tau source}

Can CP violation in the lepton sector produce the observed baryon asymmetry?  The asymmetry for the benchmark point in \cref{table:tau_source} is fairly close to the observed value in \cref{BAU}.  The scale $\Lambda_\tau \sim 1\,$TeV if fairly low, but not inconsistent with EDM and collider experiments. It could be even lowered somewhat to further increase $Y_B$ but the EFT description becomes problematic for such low scales. As shown in the right panel of \cref{fig:tau_source} the benchmark $v_w = 0.05$ is already close to optimal, and there is little room for improvement.
Definite statements about the viability of the scenario are not easy to make, as the performed calculations still suffer from uncertainties, such as those related to the description of the phase transition, that are hard to quantify. We refer to e.g. Ref.~\cite{Morrissey:2012db} for a more general discussion of  outstanding problems in calculations of EWBG. Nevertheless, our analysis indicates that CPV sources in the tau sector are viable despite the small Yukawa coupling and not yet significantly constrained by EDM experiments.

Our study of the dependence of $Y_B$ on the value of $y_\tau$ can be directly used to study the cases of CP-violating electron and muon source terms. In such cases, the transport equations are identical after replacing the third-generation lepton number densities with the first- or second-generation densities and rescaling the $y_\tau \rightarrow y_{\mu,e}$. The left panel of \cref{fig:tau_source} shows that for Yukawa couplings $y_l$ larger than $y_\tau$ we are in regime \textbf{b} where $Y_B$ grows linear with $y_l$. For smaller Yukawa couplings however, we are in regime  \textbf{a}, where $Y_B$ decreases as $y^2_l$. For equal values of the scale of new physics  $\Lambda_e= \Lambda_\mu = \Lambda_\tau$ the values of $Y_B$ in the electron and muon scenarios are then suppressed by a factor $(y_e/y_\tau)^2 \simeq 8\cdot 10^{-8}$ and $(y_\mu/y_\tau)^2 \simeq 4 \cdot 10^{-3}$, respectively. For the electron, EDM constraints require $\Lambda_e > 5.7$ TeV, which suppresses the baryon asymmetry even more. The resulting values of $Y_B$ are far below the observed asymmetries and we conclude that electron and muon CPV sources do not lead to successful baryogenesis.


\section{Baryogenesis with a quark source}
\label{sec:quarks}

In this section we discuss baryogenesis with a quark CPV source and at first neglect the additional lepton-quark interaction in section \ref{sec:GammaQL}. We start the discussion with a CPV source in the top-quark sector. This scenario has been discussed extensively in the literature, see for example Refs.~\cite{Fromme:2006wx, Huang:2015izx, Balazs:2016yvi} where the same dimension-six CPV operator has been considered. Since the source is proportional to the value of the Yukawa coupling squared $S_f \propto y_f^2$, the baryon asymmetry generated by a top source can be expected to be larger than for the corresponding source terms involving lighter fermions. But several factors that were identified in section \ref{s:toplepcompare} suppress the corresponding BAU. We therefore study how effective a top source actually is. In the process we investigate the role of SM bottom and tau Yukawa interactions that are typically neglected.

A CPV bottom source is suppressed by $\left(y_b/y_t\right)^2 \simeq 4 \times 10^{-4}$ with respect to a top source. Part of this suppression can be compensated as EDM experiments do not set as stringent constraints on the CPV bottom source. In addition, a smaller Yukawa coupling also implies less washout because $\Gamma_M^{(b)}$ is smaller and, as demonstrated for a tau source,  it is not immediately clear how $Y_B$ varies with the size of the Yukawa coupling, and thus how the top and bottom scenarios compare. While the bottom Yukawa coupling is still roughly a factor two larger than the tau Yukawa coupling, we nevertheless expect the asymmetry to be suppressed with respect to the tau source, because the bottom has strong sphaleron interactions and a smaller diffusion constant leading to less efficient generation of $Y_B$.

\subsection{Top source}
\label{s:top}

\begin{table}[t]
\renewcommand{\arraystretch}{1.2}
	\centering
	\begin{tabular}{ | l || c | c |c|c|c| }
	\hline 
          Approximation & FR$(q)$ & N$(q,t,h)$  &N$(q,t,h,u)$  &N$(q,t,h,u,l)$  &A$(q,t,h,u,l)$ \\
          \hline
  $Y_B$ & $1.6 \times 10^{-12}$ & $3.5 \times 10^{-13}$ & $3.4 \times 10^{-13}$ & $1.5 \times 10^{-12}$ & $1.1 \times 10^{-12}$     \\ \hline 
    \end{tabular}
    \caption{
      Baryon asymmetry $Y_B$ for a top source using different approximations discussed in the main text. We used the benchmark values for the bubble-wall profile given in \cref{sec:FOPS} and $\Lambda_t = 7.1$ TeV consistent with EDM experiments. 
    }   \label{table:top_source}
\end{table}

We calculate the baryon asymmetry arising from a top CPV source using various approximations: 
\begin{enumerate}
\item The simplest approximation is to first neglect the small bottom and tau Yukawa couplings.  Using \cref{ss,localbc} leaves us with three transport equations for the number densities $q$, $t$, and $h$. Finally, we apply the often-used fast-rate approximation which assumes that the top Yukawa and strong sphaleron transitions are fast and (nearly) in equilibrium ref.\cite{Huet:1995sh,Lee:2004we}. This leaves us with a single equation for $q$ that can be solved analytically. We denote the solution by FR$(q)$.
\item We neglect bottom and tau Yukawa couplings, but do not use the fast-rate approximation. Instead we numerically solve the set of three transport equations for $q$, $t$, and $h$ and call the solution N$(q,t,h)$.
\item Next we include the effects of the bottom Yukawa. We can no longer connect the $b$ density to those of light quarks and the transport equation for $u$ has to be included. The numerical solution of the four transport equations is called N$(q,t,h,u)$.
\item We also include the tau Yukawa coupling and add the $l$ transport equation. In principle, we should keep the right-handed lepton fields too, but, as discussed in the tau scenario, the approximation of local lepton number conservation,  $D_l = D_\tau$, is reasonable and we can eliminate the right-handed leptons. The solution is called N$(q,t,h,u,l)$.
\item We solve the same set of equations using the semi-analytical method \cite{White:2015bva}, which neglects the variation of the rates over the bubble wall,  and write this solution as A$(q,t,h,u,l)$.
\end{enumerate}
As before between brackets in $A(...)$ and $N(...)$  we list the transport equations that we solve explicitly.
The obtained values for $Y_B$ for our benchmark point with $\Lambda_t = 7.1 \times 10^3\,$GeV and bubble-wall velocity $v_w =0.05$ in the various approximations are listed in \cref{table:top_source}. To get the right sign for $Y_B$, the phase in the CPV operator in \cref{L6} is set to $s_t = 1$.

The fast-rate approximation assumes thermal equilibrium for the strong sphaleron and top Yukawa rates, but this approximation is invalid close to the bubble wall in the symmetric phase and invalid everywhere in the broken phase where $\Gamma_M^{(t)}$ is instead the fastest rate. By comparing the solutions FR$(q)$ and N$(q,t,h)$ we see that the fast-rate approximation overestimates $Y_B$ by roughly a factor five and is thus a poor approximation, in agreement with the findings of Ref.~\cite{White:2015bva}.  The fast-rate approximation should  be avoided to calculate $Y_B$, especially since it does not represent a systematic expansion and as such the associated error cannot be systematically estimated.

By comparing N$(q,t,h)$ and N$(q,t,h,u)$ we see that including bottom Yukawa interactions only provides a few-percent correction. Considering the significant intrinsic uncertainties associated to EWBG calculations, neglecting the bottom Yukawa seems a good approximation. It might come as a surprise then that including the even smaller  tau Yukawa interaction has a much greater impact. The baryon asymmetry associated to N$(q,t,h,u,l)$ is roughly five times that of N$(q,t,h,u)$. Where does this lepton-induced enhancement stem from?

\begin{figure}
	\centering
	\includegraphics[height = 0.35 \linewidth]{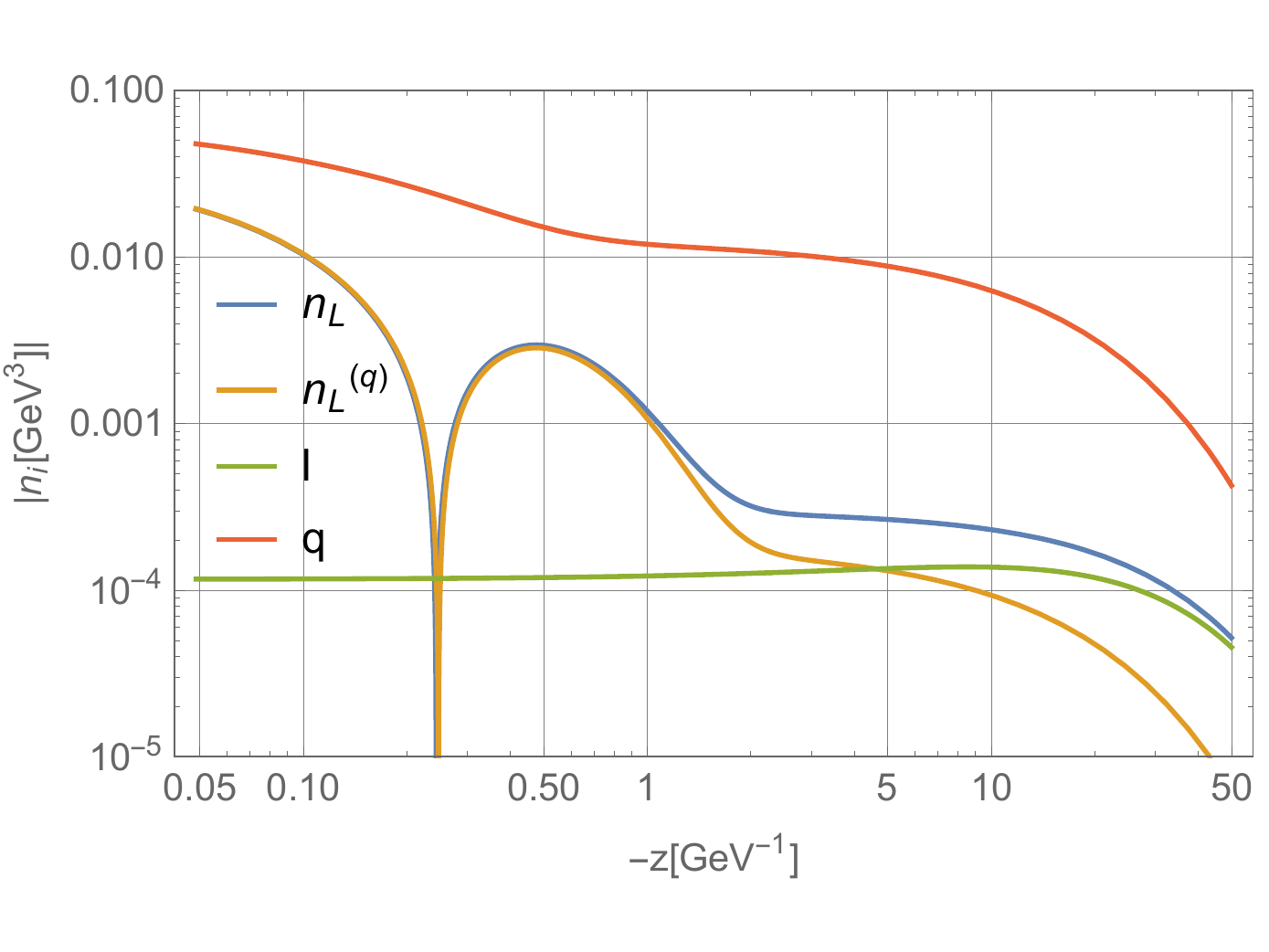}
        \includegraphics[height = 0.35 \linewidth]{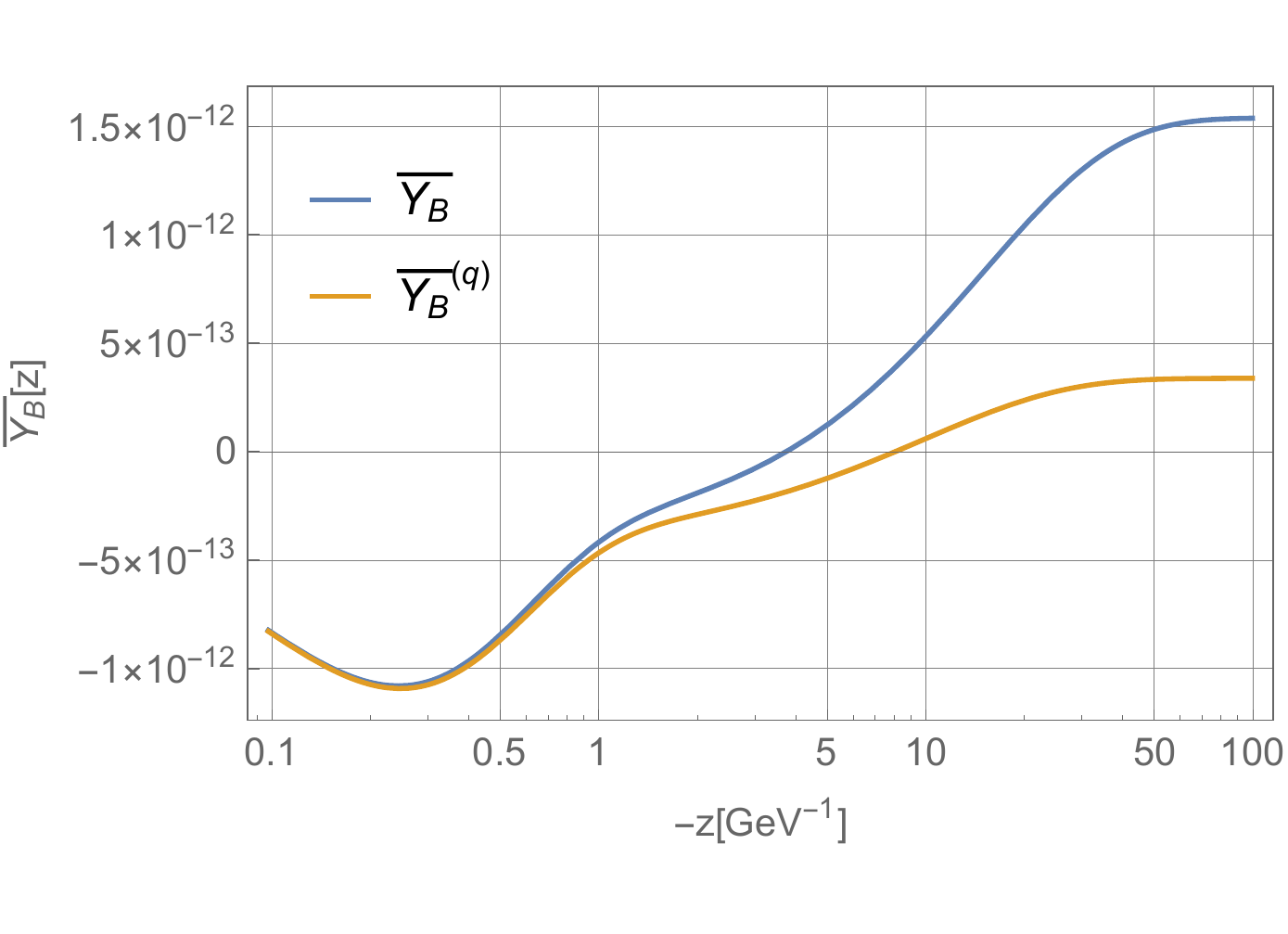}
	\caption{Left plot: absolute value of the number densities $n_L$, $n_L^{(q)}$, $l$, and $q$ in the symmetric phase for a top source. The computation includes top, bottom and tau Yukawa interactions. Right plot: baryon asymmetry $\bar Y_B(z)$ and $\bar Y^{(q)}_B(z)$ as a function of the integration cut-off.}
	\label{fig:top}
      \end{figure}

As discussed in \cref{s:toplepcompare}, when the strong sphaleron interactions are (approximately) in thermal equilibrium, $\mu_{\rm ss} \simeq 0$, the total chiral asymmetry in quarks is highly suppressed.  This is illustrated in the left panel of \cref{fig:top}. The red line depicts the chiral asymmetry in just top quarks, $q(z)$, as a function of the distance to the bubble wall in the symmetric phase. This individual asymmetry is everywhere much larger than the total chiral asymmetry in the quark sector,  $n_L^{(q)}(z) \equiv \sum_{i} q_i$, where the sum runs over the three generations. $n_L^{(q)}(z)$ is plotted in orange and is itself larger than the chiral asymmetry in tau leptons, $l(z)$, depicted in green, up to $|z|\simeq 5$. For larger distance $l(z)$ becomes larger and consequently dominates the total chiral asymmetry, $n_L(z) = n_L^{(q)}(z)+ l(z) $, depicted in blue.

The distance from the bubble in the symmetric phase at which point the various densities start to drop can be understood qualitatively from the analytical solution for the lepton source discussed in the previous section.  In analogy with the leptonic case there are three important length scales:  the diffusion length $L^i_{\rm diff} \sim D_i/v_w$, the migration length set by strong sphaleron interactions $L_S^{(\rm SS)} \sim \sqrt{k_q D_q/\Gamma_{\rm ss} }= {\mathcal O}(1) / {\rm GeV}$, and  the migration length set by the top Yukawa interactions $L_S^{(y_t)} \sim \sqrt{k_q D_q/\Gamma^{(t)}_{y} }= {\mathcal O}(0.1) / {\rm GeV}$.  The Yukawa interactions are strongest and provide the dominant source of washout near the bubble wall. This washout leads to the decrease of $n_L$ for small values of $z$ and the associated zero-crossing around $|z| = 0.2/$GeV. At distances around $|z| \sim L_S^{(y_t)}$, the top Yukawa interactions are in equilibrium and the $n_L$ would plateau to a constant value. However, this is not clear in \cref{fig:top} where $n_L$ keeps decreasing because of washout from strong sphalerons  which is active up to the larger distance $L_S^{(\rm SS)}$. This washout, and the related decrease of $n_L$, ends for $|z| > L_S^{(\rm SS)}$.

The individual particle densities migrate much further in the symmetric phase than the total quark chiral asymmetry, as their spread is only limited by diffusion.  Indeed with both the top Yukawa and strong sphaleron interactions in equilibrium, the number densities of the left- and right-handed quarks and the Higgs are related (by the condition that the chemical potential vanishes). Physically what happens is that on the relevant time scales, quarks and Higgses are converted into each other instantly, and they behave as a single degree of freedom that diffuses into the symmetric phase. Their diffusion length, which determines how far the densities extend, is dominated by the Higgs diffusion length which is large as the Higgs does not feel strong interactions. As such $L_{\rm diff} \sim D_h/v_w \simeq 20/\mathrm{GeV}$.  A more precise estimate is obtained from the transport equation for the single degree of freedom, which gives (see eq. (77a) in \cite{Lee:2004we}) $L_{\rm diff} \simeq 16/\mathrm{GeV}$.

Finally, the tau Yukawa interactions are out of equilibrium throughout.  The lepton number density slightly increases away from the bubble wall as there is more time to convert Higgs quanta into leptons. The lepton asymmetry migrates over distances set by the lepton diffusion length $L_{\rm diff} \sim D_l/v_w \simeq 20/\mathrm{GeV}$.  Because of the small tau Yukawa coupling the lepton density is always much smaller than the left-handed top density $q$. However, it is the chiral density that matters, and given the very efficient suppression of the chiral asymmetry in quarks,  leptons actually start to dominate $n_L$ for $|z| \gtrsim L_S^{(\rm SS)}$.  This explains why the inclusion of leptons can give a sizeable correction to the final baryon asymmetry. 

What is  maybe surprising is that the leptons actually give the dominant contribution for our benchmark point. The reason for this is that $n_L^{(q)}$  crosses zero close to the bubble wall around $|z| =0.25/\mathrm{GeV}$. In the right panel of \cref{fig:top} we show the contribution to the asymmetry from the wall up to a distance $z$. That is, we plot the function
\begin{equation}
  \bar Y_B(z) = 
-\frac{3  \Gamma_{\rm ws}}{2 s D_q \alpha_+} \int_{z}^0 \dd z' \, n_L(z') \e^{-\alpha_- z'}\,,
  \label{Ybz}
\end{equation}
with $\alpha_\pm $ given in \cref{alpha}.
For our benchmark parameters, the chiral asymmetry close to the bubble wall $n_L \simeq n_L^{(q)}$ gives a negative contribution to the integral  in \cref{nb_sol}. However, past the zero-crossing the contribution becomes positive. Due to the sign change, the total asymmetry generated up to $z= 10/\mathrm{GeV}$ from just $n_L^{(q)}$ vanishes. Around this point,  the chiral asymmetry in quarks is small and leptons dominate the chiral asymmetry. Integrating over larger distances leptons then give the dominant contribution  to $Y_B$. 
This sensitivity of $Y_B$ to the tau Yukawa interactions is thus twofold: 1) even though small, they are the only mechanism via which the chiral asymmetry is transferred from the quark sector, where the CPV source is located, to the lepton sector, where they escape the efficient washout by strong sphalerons. 2) There is a cancellation of the contribution of $n_L^{(q)}$ to the baryon asymmetry.

We can now also understand why including bottom Yukawa interactions has less impact. These will transfer some of the top chiral asymmetry into a bottom chiral asymmetry, but the total is still washed out by strong sphaleron interactions. This transfer therefore does not lead to a significant change in $Y_B$.

\begin{figure}
	\centering
	        \includegraphics[width = 0.5 \linewidth]{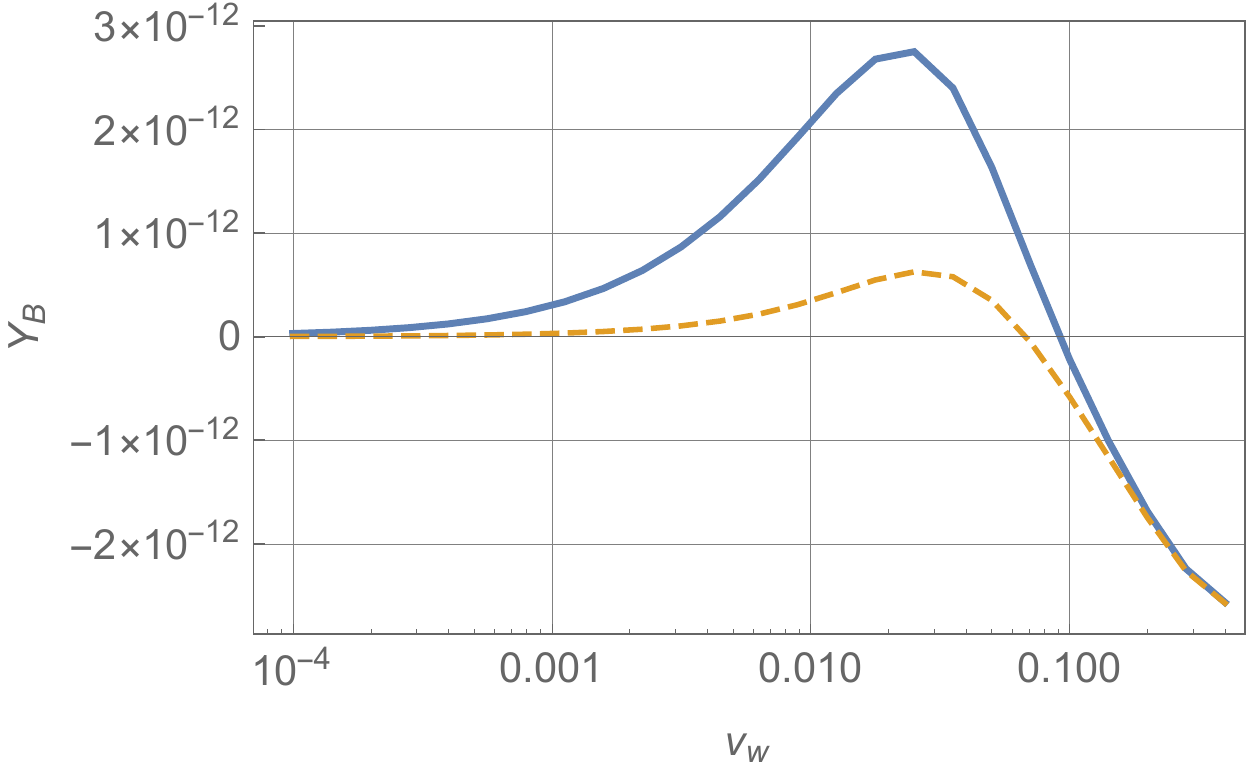}
	\caption{ $Y_B$ as a function of velocity $v_w$ with (solid blue) and without (dashed orange) lepton interaction included.}
	\label{fig:top_vw}
      \end{figure}

An interesting question is now whether including tau leptons is generally important or whether it is a special feature of our benchmark parameters. To study this, we calculate $Y_B$ associated to the N$(q,t,h,u)$ and N$(q,t,h,u,l)$ solutions for a wide range of bubble-wall velocities. The values of $Y_B$ are plotted in \cref{fig:top_vw}. 
We conclude that the lepton contribution is important for a large range of velocities $v_w \sim 10^{-4} -10^{-1}$, but becomes less relevant for larger velocities. For such large velocities the baryon asymmetry is dominated by the chiral asymmetry close to the bubble wall, where the leptons contribute little. In this case, the integral is cut off by the diffusion length $L_{\rm diff} \propto 1/v_w$, which becomes small.
     
\subsection{Producing the universal baryon asymmetry with a top source}

Let us finally answer whether a top source can produce the observed baryon asymmetry in \cref{BAU}.  For equal Wilson coefficients of the dimension-six CPV operators, $\Lambda_t = \Lambda_\tau$,  the asymmetry induced by the top source is of the same order as for the tau source. However, for taus we can choose a fairly low scale $\Lambda_\tau \sim 1\,$TeV, while EDM experiments constrain $\Lambda_t \geq 7.1$ TeV.  As such, the asymmetry in our benchmark scenario is almost two orders of magnitude too small.  As the asymmetry is rather sensitive to the Higgs vev at nucleation, this might be the easiest parameter to adjust to boost the asymmetry, but how much this is allowed depends on the details of the scalar sector that we have not specified. Our chosen bubble-wall velocity is already close to the optimal value as shown in \cref{fig:top_vw}. All things considered it seems unlikely that the dimension-six top Yukawa interaction can still lead to sufficient asymmetry.

It must be said, that the uncertainties in the calculation for a top CPV source are large. The vev-insertion approximation is dubious as the top Yukawa coupling is not small. Another potentially large effect can be the inclusion of collective plasma effects, the so-called hole modes, in the calculation of the source term and interaction rates, see the discussion in Ref.~\cite{Morrissey:2012db}. Despite these caveats, it is fair to say that the observed baryon asymmetry is obtained more easily using a tau source, mainly due to the significant EDM constraints. In \cref{s:extra} we discuss ways of boosting the baryon asymmetry for top-source scenarios by adding additional BSM lepton-quark interactions. Other ways out could be by considering CP-violating interactions between top quarks and new scalar fields as such interactions are not directly constrained by EDMs if the new scalar field does not couple to electrons. Such a setup can appear in, for example, two-Higgs doublet models although several couplings have to be set to very small values by hand.

\subsection{Bottom source}
\label{s:bottom}
\begin{table}[t]
\renewcommand{\arraystretch}{1.2}
	\centering
	\begin{tabular}{ | l || c | c |c| }
	\hline 
          Approximation  & N$(q,t,h,b)$  &N$(q,t,h,b,l)$  &A$(q,t,h,b)$ \\
          \hline
  $Y_B$ & $8.3 \times 10^{-13}$ & $8.4 \times 10^{-13}$ & $7.3 \times 10^{-13}$  \\ \hline 
    \end{tabular}
    \caption{
      Baryon asymmetry $Y_B$ for a bottom source
      with  $\Lambda_b = 1\,$TeV  and $v_w =0.05$ using different approximations discussed in the text.
    }
    \label{table:bottom}
\end{table}
We now turn to the bottom source and investigate whether such a source can be as efficient as a tau source. To calculate the chiral asymmetry generated by a bottom source, one can use the same set of transport equations as for the top source. We calculate the asymmetry first by neglecting leptons and write this solution as $N(q,t,h,u)$.  We then include tau Yukawa interactions and add the $l$-equation to the set of transport equations and obtain the solution $N(q,t,h,u,l)$. The corresponding semi-analytical solution is called A$(q,t,h,u,l)$. 
 The baryon asymmetry for our benchmark point with  $\Lambda_b = 1\,$TeV and bubble-wall velocity $v_w = 0.05$ is listed in \cref{table:bottom} (with $s_b =-1$ the sign of the dimension-six operator).

\begin{figure}[h!]
	\centering
	\includegraphics[height = 0.35 \linewidth]{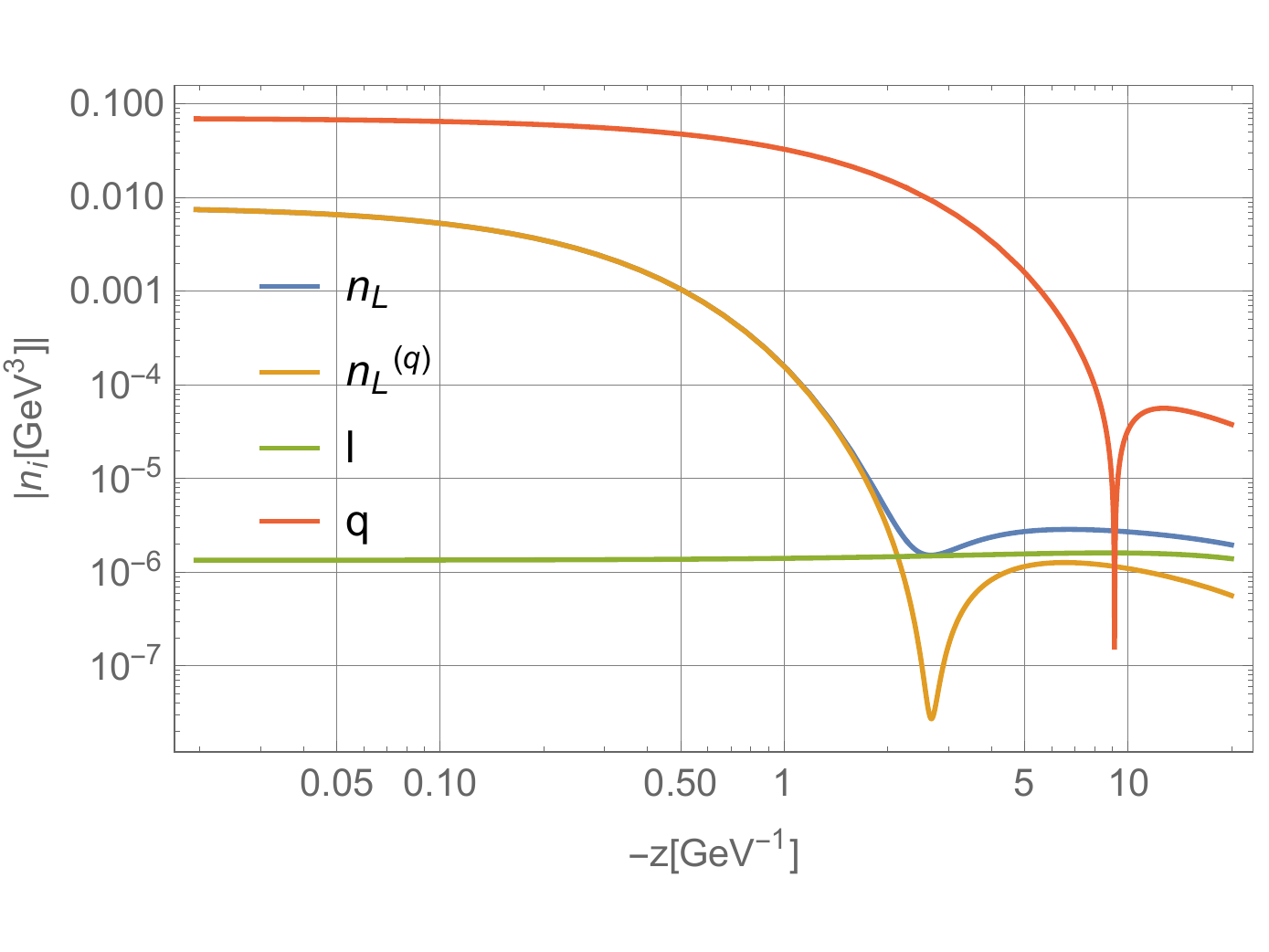}
		\includegraphics[height = 0.35 \linewidth]{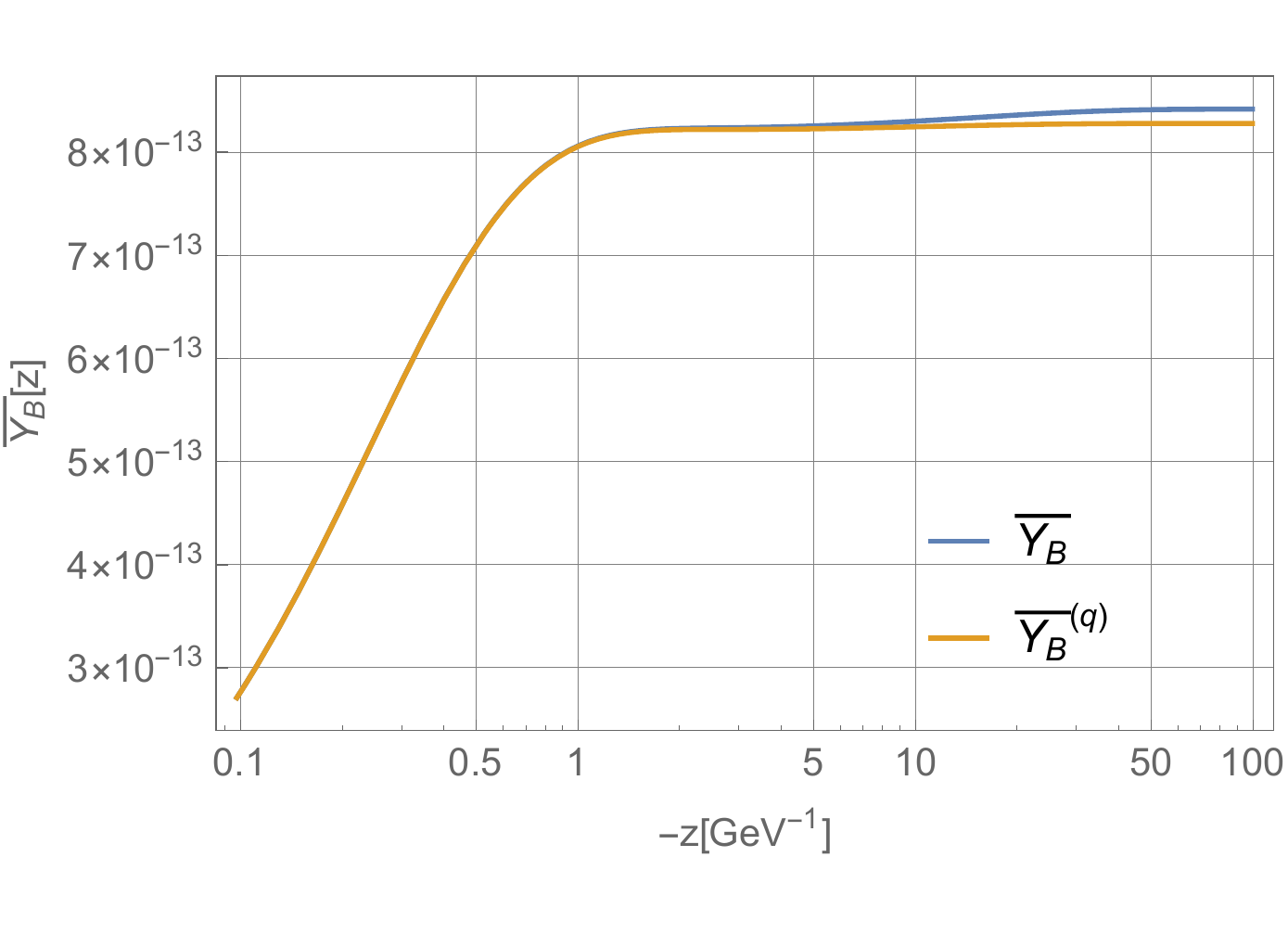}
	\caption{Left plot: absolute value of the number densities $n_L$, $n_L^{(q)} $, $l$ and  $q$ in the symmetric phase for a bottom source. The computation includes top, bottom and tau Yukawa interactions. Right plot: baryon asymmetry $\bar Y_B(z)$ and $\bar Y_B^{(q)}$ as a function of the integration cut-off.}
	\label{fig:bottom}
\end{figure}

From the table it is clear that the asymmetry from a bottom source is almost two orders of magnitude smaller than the baryon asymmetry from a tau source for the same scale $\Lambda_b=\Lambda_\tau$.  Just as for the top source, the chiral asymmetry in bottom quarks is effectively erased by the strong sphaleron interactions, which become important at $|z| \sim L_S^{\rm (SS)}$.  Beyond this scale the leptons again give the dominant contribution to the chiral asymmetry, as can be seen in the left panel in figure \ref{fig:bottom}.  However, unlike the top-source case, there is no zero-crossing, and thus no cancellation in $Y_B$. Now, the total baryon asymmetry is dominated by the contribution close to the bubble wall, where leptons are negligible, see the right panel in \cref{fig:bottom}. 

\begin{figure}
	\centering
	\includegraphics[width = 0.48 \linewidth]{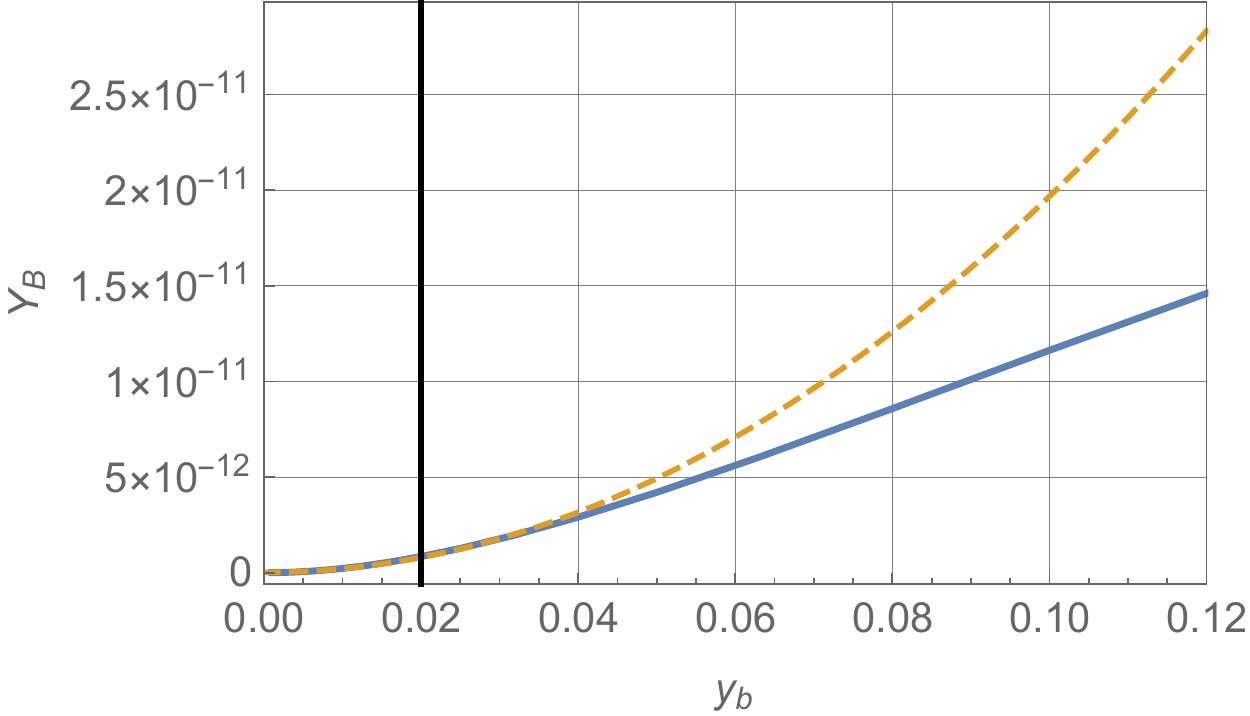}
	\caption{Baryon asymmetry for a bottom source as a function of the bottom Yukawa coupling $y_b$ (solid blue line). The orange dashed line shows a quadratic fit and the black vertical line indicates the SM value of $y_b$ at the electroweak scale. } 
	\label{fig:bottomyscaling}
\end{figure}

We conclude that the bottom source produces a value of $Y_B$ that is about two orders of magnitude too small.  What about sources involving even lighter quarks? In \cref{fig:bottomyscaling}, the solid blue line depicts the value of $Y_B$ as a function of the bottom Yukawa coupling $y_b$ at the electroweak scale. The orange dashed line is a fit to a quadratic function of $y_b$. Effectively, for the SM value of $y_b$ and smaller values we are in regime \textbf{a} discussed in \cref{sec:pardep} where $Y_B \sim y_b^2$. Since the system of transport equations for an up, down, strange, or charm source is very similar to that of a bottom source, we expect that sources involving light quarks will be suppressed by $(y_q/y_b)^2$ compared to the numbers in \cref{table:bottom}. Such sources will thus provide negligible contributions to $Y_B$. For $y_b \geq 0.04$ the asymmetry starts to scale linearly $Y_B \sim y_b$ (the system enters regime \textbf{b} discussed in \cref{sec:pardep}). The linear instead of quadratic scaling partially explains why a top source is not as effective compared to a bottom source as might be expected from just looking at the CPV source term.

\section{Consequences of additional quark-lepton interactions}
\label{s:extra}

In the previous sections we have argued that CP-violating leptonic sources are more effective than quark sources in generating a net baryon asymmetry. In this section, we analyze how we can enhance the asymmetry by considering BSM interactions that transfer the chiral asymmetry from quarks into leptons (and vice versa). Such interactions can be induced in BSM models by the exchange of new particles such as additional scalar bosons or leptoquarks.  In particular, 
we consider the effect of turning on a new top-tau coupling, which we parameterize
by the dimension-six operator in \cref{L_QL}.  

\begin{figure}
\centering
  \includegraphics[width = 0.48\linewidth]{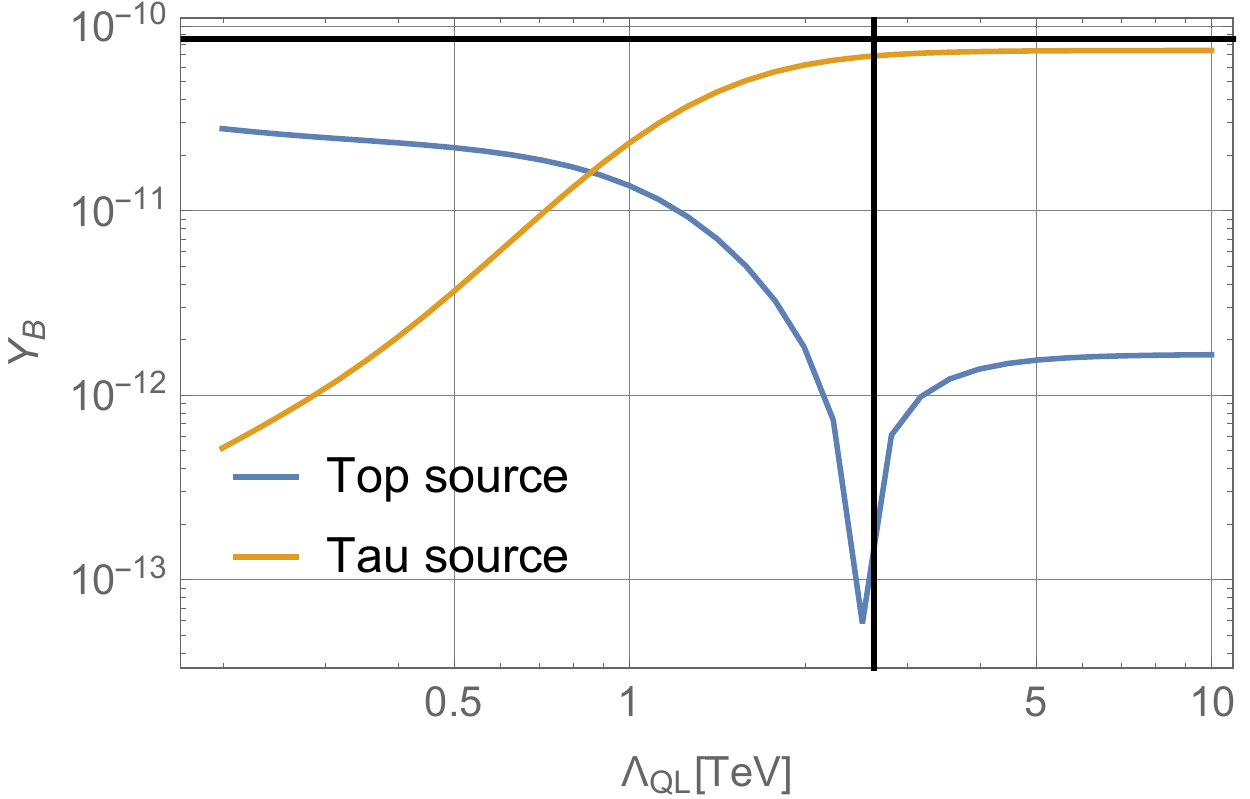}
  \caption{$Y_B$ as function of $\Lambda_{\rm QL}$ for a top source with $\Lambda_t = 7.1 \times 10^3\,$GeV (blue) and a lepton source with $\Lambda_\tau = 1 \times 10^3\,$GeV (yellow). The vertical black line corresponds to the new interactions rate being equal to the lepton Yukawa interaction rate. The horizontal black line corresponds to the measured value of the baryon asymmetry. }
  \label{fig:new} 
\end{figure}

It is intuitively clear what to expect. For a top source, the chiral asymmetry in quarks is effectively washed out by the top Yukawa and especially by the strong sphaleron interactions. In such a scenario, any part of the chiral asymmetry that is transferred to the lepton sector survives this washout, and the more effectively this is done the larger the final baryon asymmetry.  We thus expect an increase in $Y_B$ for scales $\Lambda \lesssim 3\,$TeV for which the new interaction rates becomes larger than the tau Yukawa interaction rate.  For the lepton source scenario, we expect instead a decrease in the baryon asymmetry.  The new interaction, if large, will transfer part of the chiral asymmetry from the lepton to the quark sector, where it is effectively washed out.  This behavior is indeed borne out by our numerical simulations shown in \cref{fig:new}, which shows baryon asymmetry for a lepton (yellow line) and top-quark (blue line) source, for $\Lambda_\tau = 1\,$TeV and $\Lambda_t = 7.1\,$TeV consistent with experiment.

To obtain the observed baryon asymmetry \cref{BAU} with a top source requires a large top-lepton interaction, with a fairly low scale $\Lambda_{\rm QL}\lesssim 1 \,$TeV (as always, precise statements cannot be made due to significant theoretical uncertainties).  To properly describe such a set-up probably requires going beyond the EFT description and adding the light degree of freedom that mediates the interaction. This might change the quantitative results, but we expect the results to be qualitatively the same.
As the strength of the new coupling increases, the interactions approach equilibrium, and a further increase has limited effect; this explains the asymptotic flattening of $Y_B$ for small cutoff  $\Lambda_{\rm QL}$.  The new coupling can boost the asymmetry by more than an order of magnitude. Naively one may have expected a $\mathcal O(10^2)$-increase, as the sphaleron washout reduces the chiral asymmetry in quarks by approximately this factor. However, even without the new interactions there is already  transfer taking place via the SM Yukawa interactions. While the tau Yukawa is small,  these effects already dominate the asymmetry for $\Lambda_{\rm QL}\rightarrow \infty$, as discussed in \cref{s:top}. 

While we considered here a top-tau interaction, the baryon asymmetry can be equally boosted via additional interactions between top quarks and electrons or muons as the lepton flavor is irrelevant for the baryon asymmetry. The asymmetry can also be boosted by inclusion of BSM couplings between top and one of the lighter quarks. Such couplings, if sufficiently strong, would also avoid efficient washout by strong sphalerons, as discussed in \cref{sec:GammaQL}.

The effects of the dimension-six operator in \cref{L_QL} in case of a lepton source is opposite as it now reduces the total asymmetry. As long as the new interactions are sufficiently suppressed  $\Lambda_{\rm QL}\gtrsim 1 \,$TeV, the chiral asymmetry stays mostly in the lepton sector and can be large enough to explain the observed value. However, for larger couplings the asymmetry can be suppressed by as much as two orders of magnitude. This effect can be seen in explicit models such as those studied in Refs.~\cite{Chiang:2016vgf,Guo:2016ixx} where a lepton CPV source is considered in the context of a two-Higgs doublet model. The source term, however, must be made rather large compared to the tau source discussed in \cref{s:tau} (about a factor hundred larger) to generate a baryon asymmetry consistent with observations. The source originates from CPV Yukawa-interactions with heavier non-SM Higgs fields, which can be increased without running in conflict with experiments as long as the mixing angle between the heavier Higgs and the SM Higgs is sufficiently small. EDM bounds are also avoided by choosing couplings between the non-SM scalars and electrons and lighter quarks to be sufficiently small. The heavy scalars, however, lead to effective operators of the form of \cref{L_QL} (although the actual lepton and quark flavor can vary) that, as shown in  \cref{fig:new}, suppress the baryon asymmetry. As such, larger CPV sources are required. This example shows that the rather simple framework considered here, based on effective operators can qualitatively describe the features of more complicated BSM models and can provide a useful guide in model building.

\section{Discussion and conclusions}\label{s:conclusions}
EWBG is an interesting framework for the generation of the matter-antimatter asymmetry. As EWBG takes place at a relatively low energy scale, the BSM physics required for the first-order phase transition and CPV can be tested in various experiments ranging from colliders and EDM searches to the detection of gravitational waves. EWBG however cannot take place within the SM and BSM physics is required to ensure a first-order electroweak phase transition and sufficient CPV.

Many EWBG models have been proposed and in this work we do not commit to a specific model but instead applied EFT methods to describe the dynamics of EWBG. However, Ref.~\cite{deVries:2017ncy} showed that EWBG cannot be embedded into the SM-EFT framework as a first-order phase transition requires additional light degrees of freedom that cannot be integrated out of the theory. In this work, we therefore describe the phase transition with a phenomenological ansatz and describe the required CPV with effective dimension-six operators containing SM fields only. Additional CPV operators involving the light degrees of freedom can certainly exist and be relevant, but as these are difficult to test experimentally, we leave them to future work. 

We consider flavor-diagonal CPV  dimension-six Yukawa operators of quarks and charged leptons with couplings that scale as $y_f /\Lambda_f^{2}$, where $\Lambda_f$ is the scale where new physics appears and the EFT description breaks down. The resulting CPV source term that appears in the transport equation and drives the eventual generation of the baryon asymmetry scales as $y_f^2/\Lambda_f^{2}$. As such, for the same value of $\Lambda_f$ it might be expected that the top quark would provide the largest baryon asymmetry. The recent electron EDM limit \cite{Andreev:2018ayy} sets a very strong constraint $\Lambda_t \geq 7.1$ TeV which ensures that a top CPV source as studied in this work cannot lead to the observed baryon asymmetry.  For lighter fermions the CPV source term is suppressed by $(y_f/y_t)^2$, and even though the scale $\Lambda_f$ is not as constrained, one might expect these light fermions to lead to even smaller asymmetries. We have investigated this in detail in this work and conclude that the naive scaling is not correct. Our main findings can be summarized as

\begin{itemize}
\item Despite the Yukawa suppression, a CPV tau source leads to a baryon asymmetry of the same order of magnitude as induced by a top source for $\Lambda_\tau = \Lambda_t$. As EDM constraints on $\Lambda_\tau$ are much weaker, the tau source can produce the observed baryon asymmetry where a top source cannot. The relative effectiveness of the tau source with respect to the top source has several causes. While the top source is enhanced by $(y_t/y_\tau)^2$, the Yukawa and strong sphaleron rate effectively wash out a chiral asymmetry in top quarks. This washout is far less effective for tau leptons that have much smaller Yukawa couplings and do not participate in strong sphaleron interactions. In addition, a chiral asymmetry in leptons diffuses much further into the plasma and electroweak sphaleron processes have more time to convert the chiral asymmetry into a baron asymmetry.

\item We performed analytical and numerical calculations of varying sophistication of the baryon asymmetry in case of a tau CPV source. The analytical and numerical results are found to be in very good agreement. The analytical solutions provide insight into the dependence of the baryon asymmetry on parameters related to the phase transition, such as the bubble-wall velocity, and the size of the lepton Yukawa coupling. Depending on several parameters, we can identify regions where the baryon asymmetry scales as $\propto y_f^2$, as naively expected, but also as $\propto y_f$ showing explicit deviations from the naive scaling. As such, lighter fermions can be relatively effective in generating a baryon asymmetry. 

\item While tau leptons are more efficient than might be expected, our analytical solution shows that baryon asymmetries induced by muon or electron CPV dimension-six Yukawa interactions are suppressed by respectively $(y_\mu/y_\tau)^2$ and $(y_e/y_\tau)^2$. Such sources therefore lead to much too small baryon asymmetries and only the tau source is still phenomenologically relevant.

\item Even in case of a top CPV source, leptons play an important role. The SM Yukawa interactions can convert a chiral asymmetry in quarks to a chiral asymmetry in leptons. This conversion is proportional to the tau Yukawa coupling and therefore often neglected. We find, however, that including tau leptons explicitly in the transport equations leads to a significant enhancement of the total baryon asymmetry up to an order of magnitude. This enhancement is a general feature over a wide range of bubble-wall velocities. We conclude that EWBG scenarios with CPV in the quark sector should explicitly include leptons in the transport equations. Despite this enhancement, the CPV top source from the dimension-six Yukawa interactions consistent with EDM experiments leads to a too small baryon asymmetry. This effectively rules out the minimal EWBG scenario of Ref. \cite{Huber:2006ri}, although it must be stressed that the involved theoretical uncertainties are large.

\item As a side result, we find that the fast-rate approximation which is often applied in studies of EWBG with CPV sources including top quarks, is not reliable. In our setup it significantly overestimates the total baryon asymmetry. This conclusion is in line with Ref.~\cite{White:2015bva}. We recommend to avoid its use and to instead solve the complete set of transport equations.

\item While the washout of the chiral asymmetry for a bottom CPV source is less effective than for a top CPV source, this does not compensate for the $(y_b/y_t)^2$ suppression of the CPV source. We find that a CPV source from dimension-six bottom Yukawa interactions leads to a baryon asymmetry that is approximately two orders of magnitude smaller than a top CPV source for the same value $\Lambda_b =\Lambda_t$. For values of $\Lambda_b$ consistent with EDM experiments this leads to a too small baryon asymmetry. For dimension-six Yukawa couplings for even lighter quarks the induced baryon asymmetries are even smaller.

\item The total baryon asymmetry can be significantly altered in BSM models with more effective chiral-symmetry-breaking interactions between quarks and leptons than are present in the SM. In this work, we have modeled such interactions with effective dimension-six top-tau interactions that can be induced in explicit BSM models by the exchange of new particles. Such interactions can enhance, in case of a CPV quark source, or decrease, in case of a CPV lepton source, the baryon asymmetry by orders of magnitude. This mechanism can be useful to guide model building.

\end{itemize}

As a final remark, we would like to emphasize that although we looked at a specific implementation of CP violation via dimension-six Yukawa operators, our qualitative conclusions are more general.  The importance of leptons and the related mechanism to boost the baryon asymmetry by transferring the chiral asymmetry from the quark to the lepton sector (or suppress it by doing the opposite), are independent of the details of our set-up.

\section*{Acknowledgements}
MP and JvdV are supported by the Netherlands Organization for Scientific Research (NWO). We thank Graham White and Kaori Fuyuto for helpful discussions. JvdV thanks the Amherst Center for Fundamental Interactions and the University of Massachusetts, Amherst, for the hospitality during the completion of this work.

\appendix

\renewcommand{\theequation}{\thesection.\arabic{equation}}
\numberwithin{equation}{section}

\section{Benchmark values}\label{A:BM}

In this appendix we list the values of various parameters for our benchmark scenario. 
For the values of the coupling constants at the electroweak scale $\mu= m_Z$ we use
\be g' = 0.36\,, \quad  g = 0.65\,, \quad  g_s =1.23\,, \,\quad
y_t = 1\,,\quad y_b =0.02\, ,\quad y_\tau = 0.01\,.
\ee
The diffusion constants  are taken from Refs.~\cite{Joyce:1994zn,Cline:2000nw}
\be
D_t \simeq \frac6{T}\,,\quad D_q \simeq \frac6{T}\,, \, \quad D_\tau \simeq \frac{100}{T} \, , \quad D_l \simeq \frac{380}{T}\, , \quad D_h \simeq
\frac{100}{T}\,.
\label{diffusion}
\ee
For the nucleation temperature we use 
\begin{equation}
	T_N = 88 \, \text{GeV} \, .
\end{equation}
The value of the Higgs field in the broken vacuum, the bubble-wall width and the bubble-wall speed are given by:
\begin{equation}
	v_N =152\, \text{GeV} \, , \quad L_w = 0.11 \, \text{GeV}^{-1} \, , \quad v_w = 0.05 \, . 
\end{equation}
The interaction rates are the same as in \cite{deVries:2017ncy}\footnote{The relaxation rate $\Gamma_M$ corresponds to $\Gamma_M^-$ in \cite{deVries:2017ncy}. The rate $\Gamma_M^+$ has been set to zero, which is a good approximation.} their numerical values are listed in \cref{table:rates}.
\begin{table}[t]
	\centering
	\begin{tabular}{ | l | c | c | }
		\hline 
    		& {\rm Broken phase} & {\rm Symmetric phase} \\ \hline
  		$\Gamma_{\rm ss}$ & 0.26 & 0.26 \\ \hline
  		$\Gamma_M^{(t)}$ & 104 & 0 \\ \hline
    		$\Gamma_Y^{(t)}$ & 2.7 & 2.7 \\ \hline
  		$\Gamma_M^{(b)}$ & $3.7 \cdot 10^{-2}$ & 0 \\ \hline
    		$\Gamma_Y^{(b)}$ & $1.1\cdot 10^{-3}$ & $1.1 \cdot 10^{-3}$ \\ \hline
		$\Gamma_M^{(l)}$ & $4.6 \cdot 10^{-2}$ & 0 \\ \hline
    		$\Gamma_Y^{(l)}$ & $1.0 \cdot  10^{-4}$ & $1.0 \cdot 10^{-4}$ \\ \hline  
		$  \Gamma_{\rm ws}$&0 & $4.7\cdot 10^{-4} $\\ \hline 
    \end{tabular}
    \caption{Asymptotic values of the interaction rates in the broken and symmetric phase. All rates are in $\text{GeV}$. The relaxation rates $\Gamma_M^{(f)}$ are a function of the bounce solution $\phi_b$ and vary over the bubble wall.}
    \label{table:rates}
\end{table}

\section{Weak sphalerons}
\label{s:onestep}

In this appendix we discuss the inclusion of the weak sphaleron rates in the transport equations \cref{transport} and compare the results with the two-step procedure described in \cref{s:EWS}.
We use the subscript $L$ to denote left-handed, and $l$ for leptons.

The EW sphaleron transitions involve 3 left-handed quarks and one left-handed lepton per family, changing $\Delta N_{\rm cs} =\Delta B =\Delta L =n_f$. The rate at which the baryon density $n_b$ approaches equilibrium is given by the principle of detailed balance  \cite{Joyce:1994bk,Riotto:1999yt}
\be \partial n_b
= -n_f  \Gamma_{\rm ws} (\mu_{\rm ws}+\mu_{\rm ws}^0)
\label{n_b}
\ee
with
\be
\mu_{\rm ws}=
\sum( 3\mu_{q_L} +\mu_{l_L}  )
   =\sum( 3 \frac{ n_{q_L}}{k_{q_L}} +\frac{ n_{l_L}}{k_L}  ) \simeq \frac12 \sum ( n_{q_L} + n_{l_L}) \equiv\frac12 n_L
\label{mu_ws}                              
\ee
with $n_i = k_i \mu_i$ the relation between the number density and the (rescaled) chemical potential, and with $n_L$ the chiral density. The summation is over families, and we approximated the  $k_i$-functions with their zero-temperature values.

The $\mu^0$-term takes into account the initial conditions. In the transport equations \cref{transport} the initial densities are zero before the passage of the bubble wall and such terms are absent.

\subsection{Two-step approach}

Electroweak sphalerons convert the chiral asymmetry into a baryon asymmetry. If the corresponding rate is slower than the other interaction rates it decouples from the transport equations, and we can model it as a two-step process.  The outcome of the transport equations for the fast processes is used as initial condition for the sphaleron equation, that is $\mu^0_{\rm ws}= \frac12 n_L^0$.

The strong sphalerons are fast on the relevant time-scale and are assumed to be in equilibrium, which sets $\mu_{\rm ss} = \sum (n_{q_L} -n_{u_R} - n_{d_R})=0$.  Physically, this means that any left-handed quarks produced by sphalerons are instantly divided over left- and right-handed quarks of the same generation via strong sphaleron transitions.  

The baryon asymmetry is relaxed by the  $\mu_{\rm ws}$-term in \cref{n_b}, which we want to rewrite in terms of $n_b$ for the equation to be closed.  The quark part we rewrite using
\be
n_b = n_b^L + n_b^R = 2n_b^L = \frac23 \sum n_{q_L} \, .
\label{nbquarks}
\ee
For the leptons we use that weak sphalerons conserve $B-L$. If the CPV source is in the quark sector, and the weak lepton Yukawa interactions are neglected, then leptons are only produced via weak sphalerons and the right-handed lepton densities are negligible.  Then
\be
n_b = n_l = n_l^L + n_l^R \simeq       n_l^L = \sum n_{l_L}
\label{approx_l} \, .
\ee
It would seem that this expression no longer holds if the source is in the lepton sector, or more generally, if lepton interactions are included, as it is then no longer a good approximation to neglect $\tau_R$.  However, neglecting the difference in diffusion constants for the left- and right-handed leptons,  both chiralities are produced in equal amounts by the Yukawa/relaxation/source interactions and their local densities cancel. The net lepton density then still arises from weak sphaleron transitions as assumed in \cref{approx_l}.

The chemical potential can be rewritten
\begin{align}
  \mu_{\rm ws}
 = \frac12\sum ( n_{q_L} + n_{l_L}) = \frac12\( \frac32 +1\) n_b  \equiv \frac13 {\mathcal R} n_b  \, ,                    
\end{align}
with $ {\mathcal R} = 15/4$.
Putting it all together then gives for the baryon asymmetry
\begin{align}
\partial n_b= -  \Gamma_{\rm ws} \(\frac32 n_L^0 +  {\mathcal R} n_b\) \, ,
 \label{nb2}                              
\end{align}
with $\partial n_b = v_wn_b' -D_q n_b''$.  

Inside the broken phase $n_b$ is constant, as there are no sphaleron transitions.  The broken phase becomes our universe, and this solution is what we are after.  Normalizing $Y_B(-\infty) = 0$ then gives \cref{YB}.

\subsection{One-step approach}

In the one-step approach we include the weak sphalerons in the transport equations, to calculate the baryon asymmetry in one go. In this case we can set $\mu^0_{\rm ws} =0$, as the initial densities are zero. Only left-handed particles are produced by the weak sphalerons, and their transport equation picks up an extra term.  For the third generation the transport \cref{transport} is changed to
\begin{align}
\partial_\mu q^\mu &= 
+ \Gamma_M^{(t)} \mu_M^{(t)}
+  \Gamma_M^{(b)} \mu_M^{(b)}
+  \Gamma_Y^{(t)}  \mu_Y^{(t)}
+ \Gamma_Y^{(b)}  \mu_Y^{(b)}
                     -2 \,  \Gamma_{\rm ss} \,  \mu_{\rm ss}
                     -3 \,  \Gamma_{\rm ws} \,  \mu_{\rm ws} \, ,
\nn\\
  \partial_\mu l^\mu &= 
+  \Gamma_M^{(\tau)} \mu_M^{(\tau)}
                       + \Gamma_Y^{(\tau)}  \mu_Y^{(\tau)}
                       - \,  \Gamma_{\rm ws} \,  \mu_{\rm ws}
                       -S_\tau \, ,      
 \label{transport3}
\end{align}
where for simplicity we have considered CPV in the tau sector only.

To solve the set of transport equations including the weak sphaleron rate we make some approximations. First, we neglect bottom and charm Yukawa interactions.  Second, as we have seen in the two-step description, it is a good approximation to neglect the Higgs density and set $h \simeq 0$. Third, the strong sphaleron and top Yukawa interactions are fast on the scale of the relevant leptonic interactions, and can be assumed in equilibrium, which gives $q \simeq t+b$ and $\frac43t \simeq \frac23q$, where we have approximated the $k_i$ functions with their zero-temperature values.
We define the new fields
\be
\bar q = \frac12(q+ t+b), \quad \bar l = \frac12( l-\tau), \quad \delta l = l+ \tau\,.
\ee
The usefulness of $\delta l$ is that it gives the net lepton number produced by weak sphalerons which vanishes in the limit of lepton number conservation; in the two-step process we only had the $\bar l$ equation as $\delta l=0$ and the baryons were decoupled.  The transport equations in terms of the new variables become
\begin{align}
   \partial_\mu\bar l^\mu &= 
+  \Gamma_M^{(\tau)} \mu_M^{(\tau)}
                       + \Gamma_Y^{(\tau)}  \mu_Y^{(\tau)}
                       - \frac12  \Gamma_{\rm ws} \,  \mu_{\rm ws}
                       -S_\tau \, ,     
\nn \\
  \partial_\mu \delta l^\mu  &
                               = - \,  \Gamma_{\rm ws} \,  \mu_{\rm ws}\,,
 \nn \\
  \partial_\mu \bar q^\mu &
                            =
                     - \frac{3}{2}\,  \Gamma_{\rm ws} \,  \mu_{\rm ws} \, .
\end{align}
These equations ensure that any left-handed quark number density produced by the weak sphalerons is immediately evenly spread among left- and right-handed quarks by the strong sphalerons. This explains the factor $1/2$ in the $\bar q$ equation (which is the averaged sum over left- and right-handed quarks) as compared to the coefficient of the weak sphalerons rate in the $q$-equation in \cref{transport3}.

We can further use that sphalerons conserve $B-L$ which gives the additional relation
\be
n_b = n_l \qquad \Rightarrow \qquad \delta l=  \frac23 \bar q\,,
\ee
which we use to eliminate the $\delta l$ equation.  Using $n_b =2\bar q$ then finally gives
\begin{align}
   \partial_\mu \bar l ^\mu &= 
+  \Gamma_M^{(\tau)} \mu_M^{(\tau)}
                       + \Gamma_Y^{(\tau)}  \mu_Y^{(\tau)}
                       - \frac12\,  \Gamma_{\rm ws} \,  \mu_{\rm ws}
                       -S_\tau \, ,     \nn\\
\partial_\mu n_b^\mu &=
                       - 3\,  \Gamma_{\rm ws} \,  \mu_{\rm ws} \, .
\end{align}
The chemical potentials
in terms of these variables are  
\begin{align}
 \mu_Y^{(\tau)} \simeq  \mu_M^{(\tau)}
                       & =\left( \frac{\tau}{k_\tau} - \frac{l}{k_l} \right)
                        =  \( -\frac{k_l+k_\tau}{k_l k_\tau} \bar l + \frac{k_l -k_\tau}{6k_lk_\tau}  n_b \), \nn \\
   \mu_{\rm ws} = &\frac12 \sum (q_L + l_L) = \frac12  (3\bar q +2 \delta l + l)
                          = \( \frac{\bar l}{2} + \frac{7}{6} n_b\)\,,
\end{align}
where we used $ l+\tau = \delta l =\frac13 n_b$ and $l-\tau = 2\bar l$.  The baryon asymmetry is the constant value in the broken phase, and the end result is $Y_B = n_b(z>0)/{s}$.

We have calculated the baryon asymmetry $Y_B$ for a lepton source for our benchmark parameters, and more general as a function of the bubble-wall velocity.  Over the whole range the difference between the one- and two-step calculation is at most a few percent.  We conclude that the two-step procedure is an accurate approximation, even in the regime where $\Gamma_{\rm ws} \gtrsim \Gamma_y^{(l)}$ and there is no clear separation of length scales.

\section{Numerical method}\label{A:numerics}

In this appendix we describe our method to solve the numerical transport equations in \cref{transport}. The transport equations are of the form
\be
v_w \vec x' -D \vec x'' + b \vec x + \vec s =0 \, ,
\label{eqx}
\ee 
with $\vec x= (q,t,...)^T$ the number densities, $v_w = v_w \mathbb{1}$, $D = {\rm diag}(D_q,D_t,...)$, $b$  a matrix that contains the rates which can be read from the transport equations, and $\vec s$ the source vector. We denote the number of equations, that is the dimension of $\vec x$, by $n_F$. These equations are solved with the boundary conditions that the number densities vanish far away from the bubble wall $\lim_{z=\pm \infty} \vec x =0$\footnote{Including both the tau and  lepton equation, or including the weak sphalerons in a one-step process, this boundary condition has to be generalized to $\lim_{z \infty} \vec x ={\rm const.}$, as some combination of number densities do not have interactions in the broken phase, and therefore do not necessary relax to zero. }.

In the asymptotic regions $z < z_-<0$ and $z > z_+>0$ the source vanishes and the rates become constant $\vec s=0$ and $b=$ constant, and we can solve the equations analytically.  We consider first the $z<z_-$ region of the symmetric phase.  For the asymptotic solution $\vec x_-$ we use the Ansatz 
\be
x_i = A_i \e^{\alpha z}, \quad i = 1,...,n_F \, ,
\label{xmin}
\ee
and substitute in the asymptotic form of the transport equations.  Setting $A_1 =1$, we can solve the system numerically for $A_2, ...,A_{n_F}$ and  $\alpha$. This gives $2n_F$ solutions from which we select the $n_F$ decaying solutions, that is the solutions with positive $\alpha_j$  ($\alpha_j>0$ for $j=1,..,n_F$).  The solution in the asymptotic symmetric phase is then
\be
x_i = \sum_{j=1}^{n_F} a_j  A_i (\alpha_j) \e^{\alpha_j z}, \quad z < z_- \, ,
\label{xmax}
\ee
with $a_j$ normalization constants.   Likewise the equations can be solved in the asymptotic region $z > z_+$ in the broken phase (now with the asymptotic $b$-values of the broken phase).  We denote the parameters in the broken phase by an overbar.  The asymptotic solution is\footnote{This asymptotic solution differs from that used in the semi-analytical method of \cite{White:2015bva}.  The difference is negligible if the boundary conditions are set at $z_+$ large enough (compared to the relevant length scales), such that the non-decaying terms in the semi-analytical solution are also negligible.  This can always be done in the planar wall approximation for the bubble profile used in this paper, but not neccessarily for the actual profile of a finite-sized bubble. }
\be
x_i = \sum_{j=1}^{n_F} \bar a_j \bar A_i (\bar \alpha_j) \e^{- \bar\alpha_j z}, \quad z > z_+ \, ,
\label{BC_broken}
\ee
with $\bar \alpha_j >0$.

The idea now is to solve the transport equations and match to the asymptotic solutions. That is, we solve the equations with boundary conditions set at $z_\pm$.  \Cref{xmin} gives $2n_F$ boundary conditions $\vec x(z_-), \, \vec x'(z_-)$ at $z_-$ in terms of $n_F$ normalization constants $a_j$, and \cref{xmax} gives $2n_F$ boundary conditions $\vec x(z_+), \, \vec x'(z_+)$ at $z_+$ in terms of $n_F$ normalization constants $\bar a_j$. The total of $2n_F$ normalization constants are determined by a shooting method.

To be more specific, we start with the analytical solution as a boundary condition at $z_-$, we pick the $n_F$ normalization constants $a_j$, and numerically evolve to $z=0$ using the full transport equations \cref{eqx}.  ce Call this solution $\vec x(0_-) \equiv \vec x_-(a_j)$ .  Likewise we pick values for $\bar a_j$, and numerically evolve the solution at $z_+$ back to $z=0$, and call this solution $\vec x(0_+) \equiv \vec x_+(\bar a_j)$.  We adjust $a_j, \bar a_k$ via a shooting procedure such that at $z=0$ the first and second derivative of the two solutions match, i.e. 
\be
\vec  x_-(a_j) =  \vec x_+(\bar a_j), \quad\vec  x'_-(a_j) =  \vec x'_+(\bar a_j)\,.
\label{cont_eq}
\ee

The shooting procedure can be implemented using Newton's method by starting with initial guesses for $a_j, \bar a_i$ and then calculating $x_-,x_+$.  As a measure of goodness of the solution we define the function 
\be
\vec F = 
\( \begin{array}{c}
 \vec x_- -\vec x_+ \\
 \vec x'_- - \vec x'_+ 
\end{array}
\)\,.
\ee
Newton's method gives a way to find the next values of  $\vec b = \binom{\vec a  }{\vec{\bar a}}$, that minimizes the $\vec F$ function (both $\vec b$ and $ \vec F$ are  $2n_F$ dimensional vectors)
\be
\vec b_{\rm new} = \vec b - J(\vec b)^{-1} \vec F(\vec b), \quad {\rm with} \quad
J = \left[
\begin{array}{c cc c}
\frac{\partial F_1}{\partial b_1} & \frac{\partial F_1}{\partial b_2}&  \cdots &\frac{\partial F_1}{\partial b_{2n_F}} \\
\frac{\partial F_2}{\partial b_1} & \frac{\partial F_2}{\partial b_2}&  \cdots &\frac{\partial F_2}{\partial b_{2n_F}} \\
\vdots & \cdots & \cdots & \vdots \\
\frac{\partial F_{2n_F}}{\partial b_1} & \frac{\partial F_{2n_F}}{\partial b_2}&  \cdots &\frac{\partial F_{2n_F}}{\partial b_{2n_F}}
\end{array}
\right]\,.
\label{jacobian}
\ee
We reiterate the process until  $\sum_i^{2n_F}  |F_i|< \eps$ smaller than the required precision.  We choose $\eps =10^{-6}$, and have checked that the result does not change for smaller values of $\eps$.

We calculate the entries of the Jacobian as follows.  We choose $\tilde b = (b_1, ..., b_i (1+ \eps), ..., b_{2n_F})$ and calculate with these entries $(\tilde x_-, \tilde x_+)$ and determine $\tilde F_j$.  Then $\partial F_j/\partial b_i = (\tilde F_j - F_j)/(b_i \eps)$. We do this for $i =1,2,...,2n_F$ to find all entries of the Jacobian.


\bibliographystyle{h-physrev3} 
\bibliography{myrefs}

\end{document}